\documentclass{emulateapj}
\usepackage{natbib}

\usepackage[latin1]{inputenc}
\usepackage{multirow}
\usepackage{graphicx}
\usepackage{float}
\usepackage{color}
\usepackage{footnote}
\usepackage{amssymb}
\usepackage[figuresright]{rotating}
\usepackage{amsmath}
\usepackage{url}
\usepackage{rotating}
\usepackage{color}
\usepackage{soul}

\newcommand{\apgt}{{\raise-.5ex\hbox{$\buildrel>\over\sim$}}}
\newcommand{\aplt}{{\raise-.5ex\hbox{$\buildrel<\over\sim$}}} 
\newcommand{\Mpc}{\rm\; Mpc}
\newcommand{\kpc}{\rm\; kpc}

\newcommand{\km}{\rm\; km}

\newcommand{\s}{\rm\; s}

\newcommand{\keV}{\rm\; keV}

\newcommand{\erg}{\rm\; erg}

\newcommand{\ergps}{\hbox{$\erg\s^{-1}\,$}}

\newcommand{\kmps}{\hbox{$\km\s^{-1}\,$}}

\newcommand{\kmpspMpc}{\hbox{$\kmps\Mpc^{-1}\,$}}

\newcommand{\Omm}{\hbox{$\rm\thinspace \Omega_{m}$}}
\newcommand{\OmL}{\hbox{$\rm\thinspace \Omega_{\Lambda}$}}

\def\aap{A\&A}                   

\def\apjs{ApJS}                  
\def\apjl{ApJ}                   

\newcommand{\myemail}{juliehl@stanford.edu}

\slugcomment{Accepted for publication in ApJ}

\shorttitle{AGN feedback in RX~J1532.9+3021}
\shortauthors{Hlavacek-Larrondo et al.}

\begin{document}


\title{Probing the extreme realm of AGN feedback in the massive galaxy cluster, RX~J1532.9+3021}


\author{J. Hlavacek-Larrondo\altaffilmark{$\star$,1,2}, S. W. Allen\altaffilmark{1,2,3}, G. B. Taylor\altaffilmark{4,5}, A. C. Fabian\altaffilmark{6}, \\R. E. A. Canning\altaffilmark{1,2}, N. Werner\altaffilmark{1,2}, J. S. Sanders\altaffilmark{7}, C. K. Grimes\altaffilmark{4,5}, \\S. Ehlert\altaffilmark{1,2} and A. von der Linden\altaffilmark{1,2,8}}
\altaffiltext{$\star$}{Einstein fellow, Email: \myemail}
\altaffiltext{1}{Kavli Institute for Particle Astrophysics and Cosmology, Stanford University, 452 Lomita Mall, Stanford, CA 94305-4085, USA}
\altaffiltext{2}{Department of Physics, Stanford University, 452 Lomita Mall, Stanford, CA 94305-4085, USA}
\altaffiltext{3}{SLAC National Accelerator Laboratory, 2575 Sand Hill Road, Menlo Park, CA 94025, USA}
\altaffiltext{4}{Department of Physics and Astronomy, University of New-Mexico, Albuquerque, NM 87131, USA}
\altaffiltext{5}{National Radio Astronomy Observatory, Socorro, NM 87801, USA}
\altaffiltext{6}{Institute of Astronomy, University of Cambridge, Madingley Road, Cambridge CB3 0HA}
\altaffiltext{7}{Max-Planck-Institut fur extraterrestrische Physik (MPE), Giessenbachstrasse, 85748 Garching, Germany}
\altaffiltext{8}{Dark Cosmology Centre, Niels Bohr Institute, University of Copenhagen, Juliane Maries Vej 30, 2100 Copenhagen, Denmark}

\begin{abstract}
We present a detailed \textit{Chandra}, \textit{XMM-Newton}, \textit{VLA} and \textit{HST} analysis of one of the strongest cool core clusters known, RX J1532.9+3021 (z=0.3613). Using new, deep 90 ks \textit{Chandra} observations, we confirm the presence of a western X-ray cavity or bubble, and report on a newly discovered eastern X-ray cavity. The total mechanical power associated with these AGN-driven outflows is $(22\pm9){\times}10^{44}\ergps$, and is sufficient to offset the cooling, indicating that AGN feedback still provides a viable solution to the cooling flow problem even in the strongest cool core clusters. Based on the distribution of the optical filaments, as well as a jet-like structure seen in the 325 MHz VLA radio map, we suggest that the cluster harbours older outflows along the north to south direction. The jet of the central AGN is therefore either precessing or sloshing-induced motions have caused the outflows to change directions. There are also hints of an X-ray depression to the north aligned with the 325 MHz jet-like structure, which might represent the highest redshift ghost cavity discovered to date. We further find evidence of a cold front ($r\approx65$ kpc) that coincides with the outermost edge of the western X-ray cavity and the edge of the radio mini-halo. The common location of the cold front with the edge of the radio mini-halo supports the idea that the latter originates from electrons being reaccelerated due to sloshing induced turbulence. Alternatively, its coexistence with the edge of the X-ray cavity may be due to cool gas being dragged out by the outburst. We confirm that the central AGN is highly sub-Eddington and conclude that a $>10^{10}M_{\rm \odot}$ or a rapidly spinning black hole is favored to explain both the radiative-inefficiency of the AGN and the powerful X-ray cavities.
\end{abstract}





\section{Introduction}
Active Galactic Nucleus (AGN) feedback plays a fundamental role in shaping the properties of massive galaxies, by injecting energy into the surrounding medium through radiation, winds and jets. 

In isolated galaxies, this energy easily escapes, leaving almost no trace due to the low densities of the surrounding gas. However, the intracluster gas in galaxy clusters is dense enough to retain an imprint of the feedback, mostly in the form of regions of reduced X-ray surface brightness known as X-ray cavities or bubbles. 

These regions originate from supersonic jets that drill through and push aside the surrounding X-ray emitting gas. They are often seen at the centres of galaxy clusters, where the jets from the supermassive black holes of the Brightest Cluster Galaxies (BCGs) are inflating them, but have also been seen in other cluster members such as NGC 4649 in the Virgo galaxy cluster \citep{Shu2008383}, as well as at the centres of groups of galaxies, where the supermassive black holes of the Brightest Group Galaxies (BGGs) are inflating them \citep[e.g.][]{Ran2011726}. Critically, X-ray cavities provide unique opportunities to quantify the energies associated with AGN feedback, by enabling straightforward calculations of the total energy required to inflate them. They therefore provide vital tools in assessing the role AGN feedback plays in the formation and evolution of massive galaxies.

\begin{figure*}
\centering
\begin{minipage}[c]{0.50\linewidth}
\centering \includegraphics[width=\linewidth]{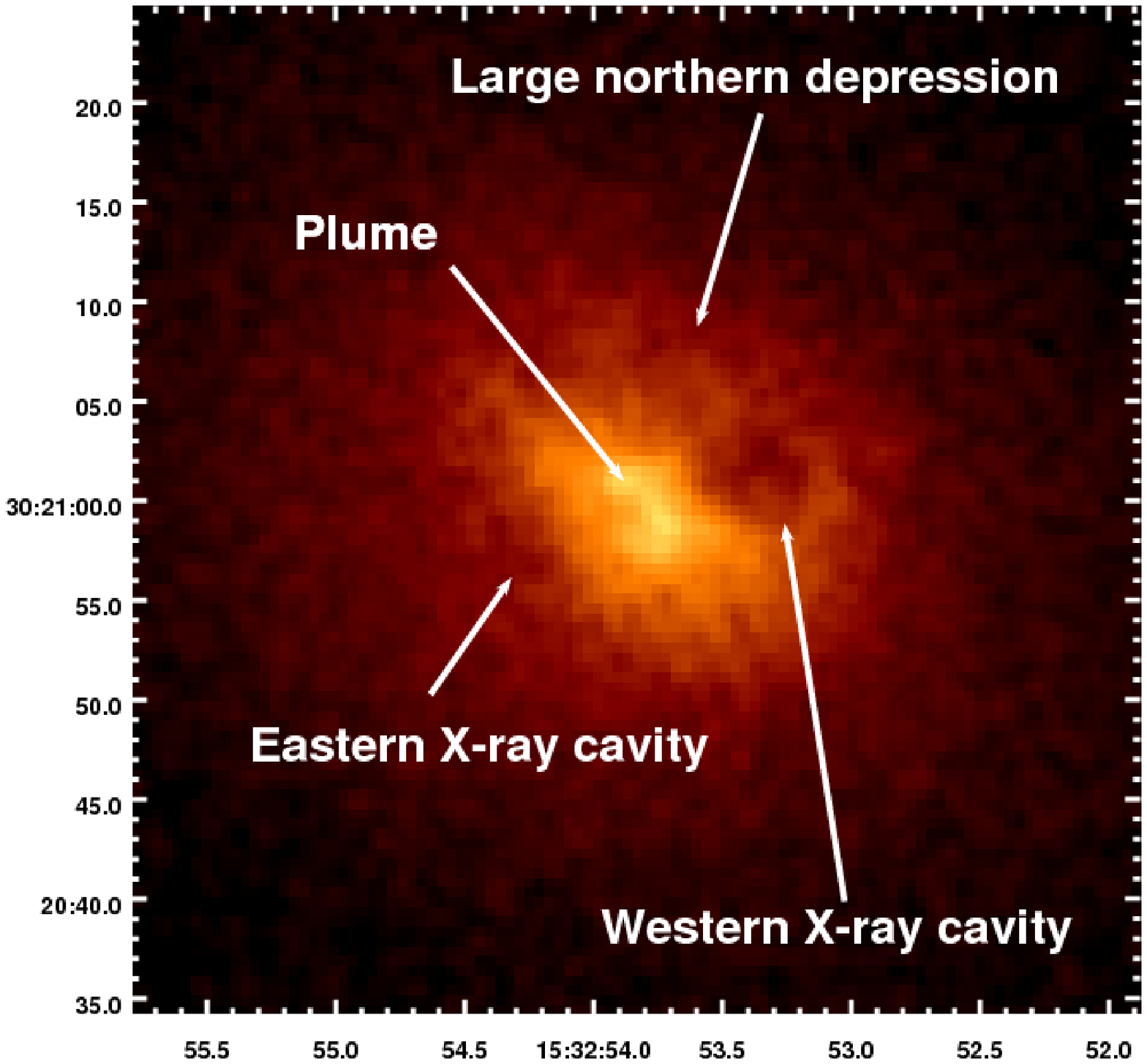}
\end{minipage}
\begin{minipage}[c]{0.485\linewidth}
\centering \includegraphics[width=\linewidth]{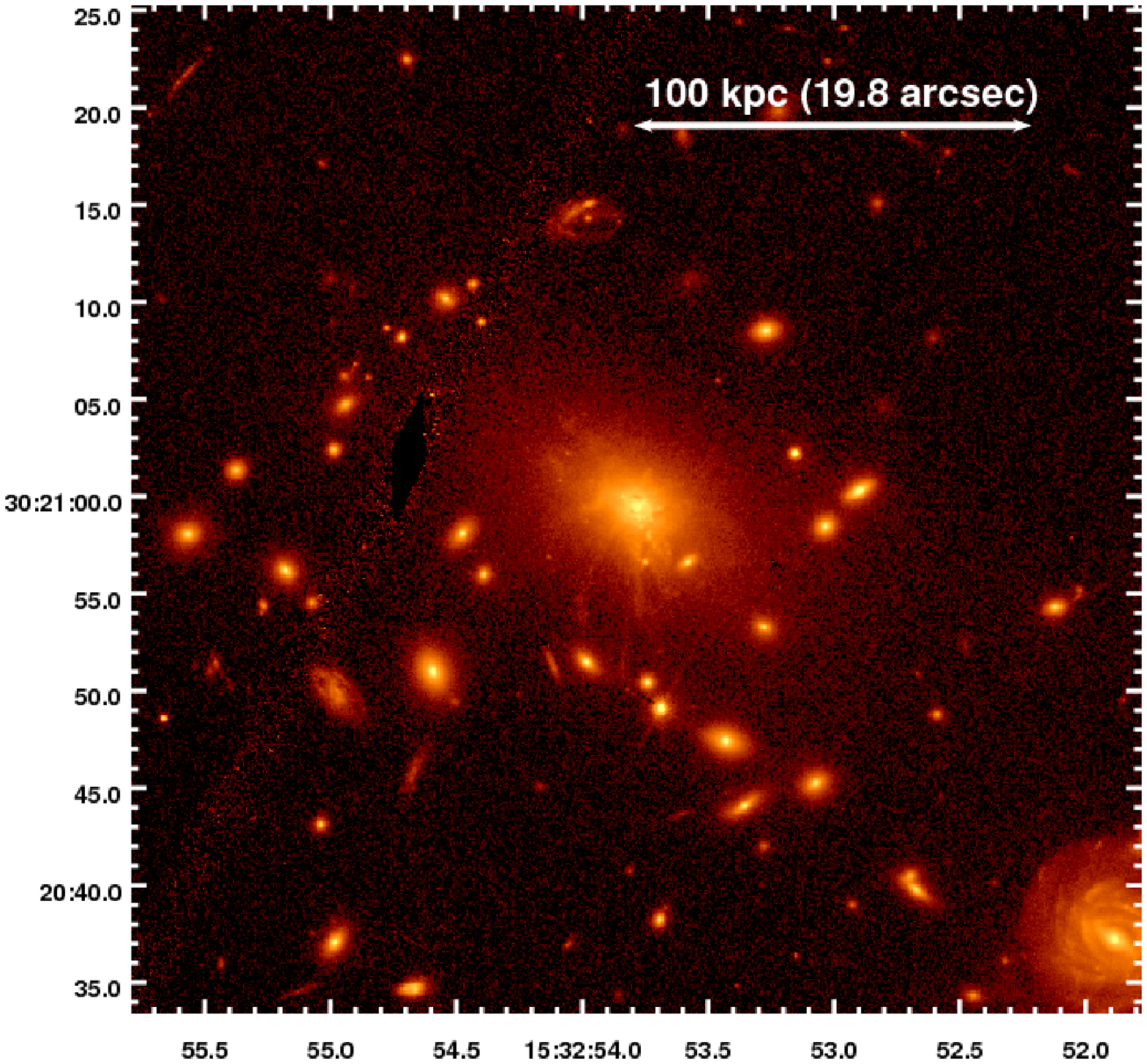}
\end{minipage}
\caption[]{Left - Exposure-corrected $0.5-7\keV$ \textit{Chandra} image of RX~J1532.9+3021 totalling 105 ks, spatially smoothed with a $0.8''$ gaussian function and covering the central $250\times250\kpc$ (or $50''\times50''$). Shown in this image are the inner two X-ray cavities, the central plume-like structure and the outer depression (potential ghost cavity) to the north. Right - Same-scale HST ACS/WFC F814W image centred on the BCG. North is up and east is to the left.}
\label{fig1}
\end{figure*}

In clusters of galaxies, studies have shown that X-ray cavities are predominantly found in systems that have steeply rising X-ray surface brightness profiles, drops in temperature by a factor of 3 near the centre and short radiative cooling times \citep[e.g.]{Edg1992258}. These are known as cool core clusters and large cooling flows at the rates of $100-1000~{\rm M_{\rm \odot}}$ yr$^{-1}$ would be expected in their centres in the absence of a balancing heating source. They account for some 50 per cent of local X-ray selected clusters \citep[e.g.][]{All1997286}.

Although some cooling is observed in cool core clusters, in the form of warm and cold molecular gas as H$_\mathrm{2}$ and CO lines \citep[][and references therein]{Fal1998494,Edg2001328} and optical and UV emission lines providing evidence for some on-going star formation \citep[e.g.][]{Joh1987224}, the measured rates of cooling fall well below the expected rates ($\aplt5$ per cent). In particular, \textit{Chandra} and \textit{XMM-Newton} have shown that there is much less gas cooling below a factor of $1/3$ of the outer temperature than is predicted from cooling models \citep{Pet2006427}. Some form of feedback in cool core clusters is therefore needed to counterbalance the cooling, and the central AGN is likely to be the energy source by inflating the X-ray cavities, inducing weak shocks and propagating energy through sound waves \citep[][]{Bir2004607,Bir2008686,Dun2006373,Dun2008385,Nul2009,Cav2010720,Don2010712,Hla2012421}. However, how the fine tuning between heating and cooling is accomplished still remains poorly understood.

To further test our understanding of AGN feedback, we have targeted the most extreme cool core clusters, requiring at least $10^{45}\ergps$ of mechanical feedback from their central AGN to prevent the intracluster medium (ICM) from cooling. Only a few examples of such extreme objects are know so far \citep{Ehl2011411,Cav2011732,McD2012Nat,Hla2012421,Ogr2010406}, one of which is the focus of this study, RX~J1532.9+3021 (MACS~J1532.9+3021).

The objective of this paper is therefore to conduct a detailed analysis of RX~J1532.9+3021, and test the fine balance between heating and cooling in cool core clusters at the extreme level. Section 2 presents the X-ray, radio and optical observations used throughout this study. Section 3 and Section 4 then present the imaging analyses, and Section 5 presents the thermodynamic properties of RX~J1532.9+3021. Section 6 focuses on the radiative and mechanical properties of the AGN, and finally, we discuss our results in Section 7. We adopt $H_\mathrm{0}=70\kmpspMpc$ with $\Omm=0.3$, $\OmL=0.7$ throughout this paper, and a cluster redshift of $z=0.3613$ \citep[$1''=5.042$ kpc;][]{Ebe2010407}. All errors are $1\sigma$ unless otherwise noted.

\section{Data reduction and processing}

\subsection{X-rays: \textit{Chandra} observations}

RX~J1532.9+3021 was originally observed in 2001 with the \textit{Chandra} CCD Imaging Spectrometer (ACIS) in VFAINT mode and centred on the ACIS-S3 back-illuminated chip (10 ks; ObsID 1649) and the ACIS-I3 front-illuminated chip (10 ks; ObsID 1665). It was subsequently observed on 2011 November 16 for $90$ ks (ObsID 14009; PI Hlavacek-Larrondo) in VFAINT mode, significantly improving the image quality and centred on the ACIS-S3 back-illuminated chip with the ACIS-S1, ACIS-S2, ACIS-S4 and ACIS-I3 also switched on. 

All three ObsIDs were processed, cleaned and calibrated using the latest version of the {\sc ciao} software ({\sc ciaov4.5}, {\sc caldb 4.5.6}), and starting from the level 1 event file. We applied both charge transfer inefficiency and time-dependent gain corrections, as well as removed flares using the {\sc {lc$\_$clean}} script with a $3\sigma$ threshold. The net exposure times are shown in Table \ref{tab1}. We then exposure-corrected the image, using an exposure map generated with a monoenergetic distribution of source photons at 1.5{\keV}. The combined exposure-corrected $0.5-7.0\keV$ image is shown in Fig. \ref{fig1}.

\begin{table}
\caption[]{\textit{Chandra} observations.}
\begin{tabular}{lccc}
\hline
\hline
Observation number & Date & Detector & Exposure (ks) \\
\hline
1649 & 	08/26/2001 & ACIS-S & 9.5 \\
1665 &  09/06/2001 & ACIS-I & 8.8 \\
14009 & 11/16/2011 & ACIS-S & 86.7 \\
\hline
\end{tabular}
\label{tab1}
\end{table}

\begin{table}
\caption[]{\textit{XMM-Newton} RGS observations.}
\hspace{-0.1in}
\resizebox{8.8cm}{!} {
\begin{tabular}{lcc}
\hline
\hline
Observation number & Date & Exposure (ks) \\
\hline
0039340101 & 2002-07-08 & ... \\
0651240101 & 2010-08-04 & 84.8 (combined) \\
\hline
\end{tabular}}
\label{tab1b}
\end{table}

\begin{table*}
\caption[]{\textit{VLA} radio observations.}
\resizebox{18cm}{!} {
\begin{tabular}{lccccc}
\hline
Date & Frequency (MHz) & Bandwidth (MHz) & Configuration & Duration (min) & rms (mJy/beam)\\
\hline
15-Mar-2006	& 	325	& 	10.2	& 	A	& 	67	& 		0.6 \\
15-Mar-2006	&	1400 &	100		&	A	&	46	&		0.023 (combined A+B) \\
11-Sep-2006	& 1400	& 100		&	B	&	43	&		0.023 (combined A+B) \\
\hline
\end{tabular}}
\label{tab2}
\end{table*}

\begin{figure*}
\centering
\begin{minipage}[c]{0.99\linewidth}
\centering \includegraphics[width=\linewidth]{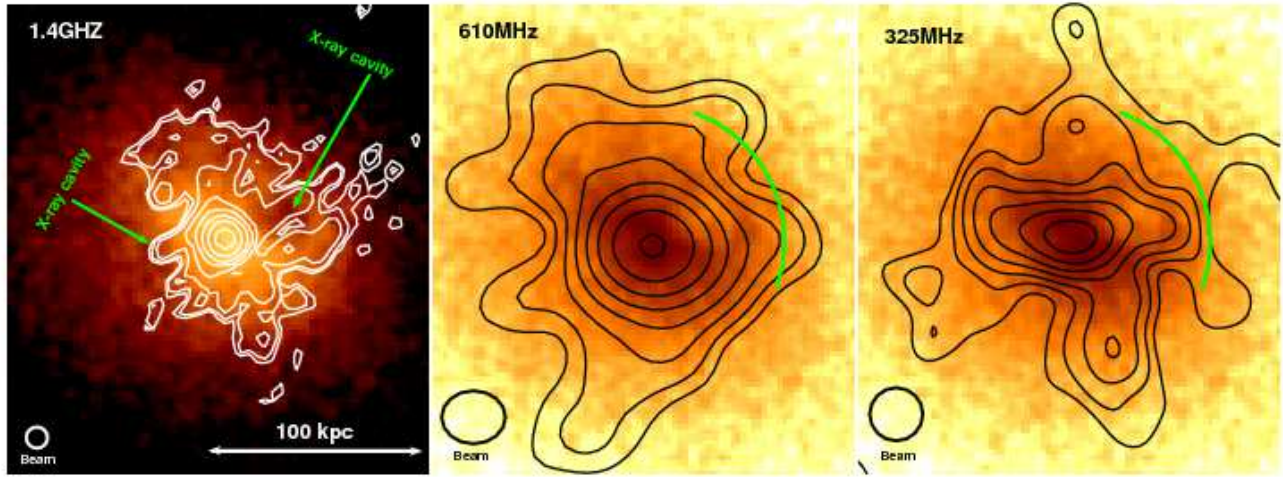}
\end{minipage}
\caption[]{Exposure-corrected $0.5-7\keV$ \textit{Chandra} image, same as in Fig. \ref{fig1}, with the VLA 1.4 GHz contours (left), \textit{GMRT} 610 MHz radio contours (middle) from \citet{Kal2013} and VLA 325 MHz contours (right). Contours start at $3\sigma_{\rm rms}$. Also shown is the beam size in the bottom-left, as well as the location of the X-ray cavities and the cold front with the green arc (see Section 7.1). }
\label{fig1b}
\end{figure*}

The X-ray data were spectroscopically analyzed with {\sc xspec} (v12.8.0a). We use the abundance ratios of \citet{And198953} throughout this study, and exclude points sources from all regions by identifying them by eye. The latter are only present beyond a radius of 250 kpc. We also mostly use C-statistics throughout the study. The background for each ObsID is selected as a region located on the same chip, devoid of point sources, and far from any cluster emission. It is also chosen to occupy the same location on the Sky in all three ObsIDs. Since RX~J1532.9+3021 has a high redshift, the cluster occupies only a small fraction of the chips, about 50 per cent within a radius of 1 Mpc, and a background can easily be extracted from the surrounding region. Note that our choice of background does not significantly affect our results. If we use blank-sky observations or select a similar region, located on the second back-illuminated chip (ACIS-S1) for ObsID1649 and ObsID14009, as well as another chip for ObsID1665 (ACIS-I0), we find results consistent with each other. 

We use the \citet{Kal2005440} value for the Galactic absorption throughout this study. If we select a region located between 50 kpc and 200 kpc, and fit an absorbed {\sc mekal} plasma model to the $0.5-7$ keV spectrum while letting the Galactic absorption, temperature, abundance and normalization parameters free to vary, we find that the derived absorption value is consistent with the \citet{Kal2005440} value to within 1$\sigma$, justifying our choice for the Galactic absorption.

\subsection{X-rays: \textit{XMM-Newton} observations}

There are two datasets available in the \textit{XMM-Newton} archive for RX~J1532.9+3021, see Table \ref{tab1b}. We only consider the high spectral resolution XMM-Newton reflection grating spectrometer (RGS) data since these provide better constraints on the amount of cool gas (see Peterson \& Fabian 2006 and references therein)\nocite{Pet2006427}. Data reduction was accomplished following the standard {\sc sas} procedures. A PSF extraction region of 90 per cent and a pulse-height distribution region of 95 per cent were used. The data were also grouped to have at least 25 counts per spectral bin and background model spectra were created with {\sc rgsbkgmodel}.

\subsection{Radio observations}
There are several observations of RX~J1532.9+3021 in the Very Large Array (\textit{VLA}) archive. We only consider the data in A and B configurations, since the spatial resolution is well matched to that of the \textit{Chandra} data. The data are summarized in Table \ref{tab2}. We reduced them using the standard astronomical image processing system ({\sc aips}, Greisen 2003)\nocite{Gre2003285}.

The combined A and B configuration 1.4 GHz image, with a 2$''$ spatial resolution is shown in the left panel of Fig. \ref{fig1b}. The best VLA 325 MHz image at 5$''$ resolution is shown in the right panel of Fig. \ref{fig1b}. The 1.4 GHz map reveals a strong core and twin elongated lobes to the east and west that coincide with the X-ray cavities. These lobes extend to the edge of the cavities along the jet direction, but are thinner along the perpendicular direction. The 325 MHz map reveals a jet-like feature along the north to south direction. We further discuss this feature in Section 7.2 in the context of the presence of older outflows, known as ghost cavities. 

There is also a faint underlying component seen in the maps that extends out to 70 kpc in radius ($\mu$Jy level). We interpret this component as a mini-halo, similar to what is seen in other cool core clusters \citep[e.g.][]{Maz2008675}. This mini-halo has also been recently reported by Giacintucci et al. (2013)\nocite{Gia2013}, and we show in the middle panel of Fig. \ref{fig1b} their Giant Metrewave Radio Telescope (\textit{GMRT}) 610 MHz radio contours \citep[which will also appear in][]{Kal2013}. 

\begin{figure}
\centering
\begin{minipage}[c]{0.8\linewidth}
\centering \includegraphics[width=\linewidth]{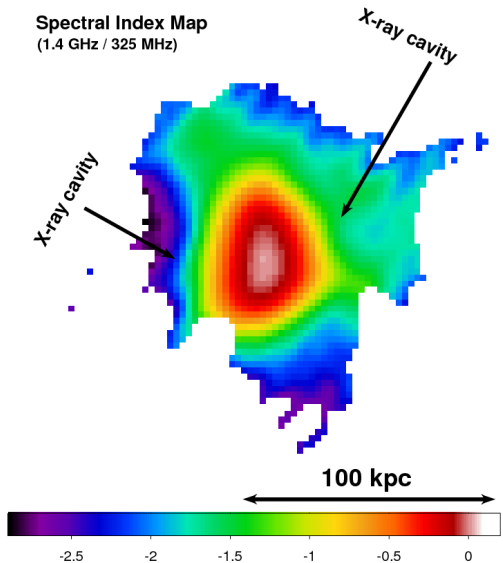}
\end{minipage}
\caption[]{Spectral index map ($\alpha$) of the radio emission computed between the VLA 1.4 GHz and VLA 325 MHz maps at a resolution of $\approx5''$. Same scale as Fig. \ref{fig1b}.}
\label{fig1c}
\end{figure}

Fig. \ref{fig1c} shows the spectral index map, obtained by calculating the index between the VLA 1.4 GHz and 325 MHz images. This map was produced by first convolving the two images to a similar resolution of $\approx5''$, and then computing the spectral index for regions where the radio emission was above $3\sigma_{\rm rms}$. The spectral index ($\alpha$) is defined such that the flux density scales as $S_\nu\propto\nu^{-\alpha}$. Interestingly, Fig. \ref{fig1c} shows that the core has a flat, and even slightly inverted radio spectrum. The extended emission on the other hand is characterized by a steep spectrum with $\alpha\approx-2$, which is typically seen in mini-halos and indicates aging of the relativistic particles producing the radio emission \citep[e.g.][]{Fer2008134}.

\begin{figure*}
\centering
\begin{minipage}[c]{0.32\linewidth}
\centering \includegraphics[width=\linewidth]{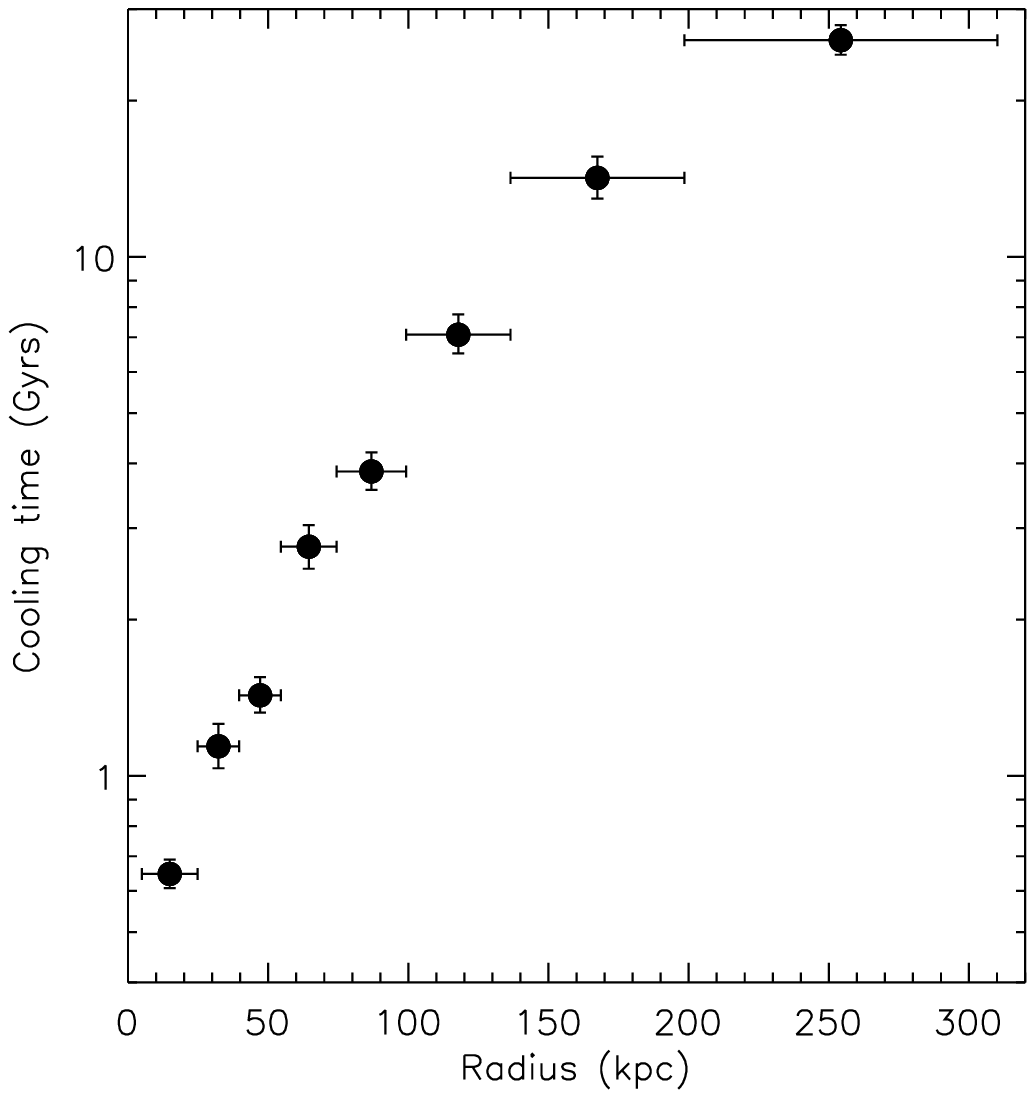}
\end{minipage}
\begin{minipage}[c]{0.32\linewidth}
\centering \includegraphics[width=\linewidth]{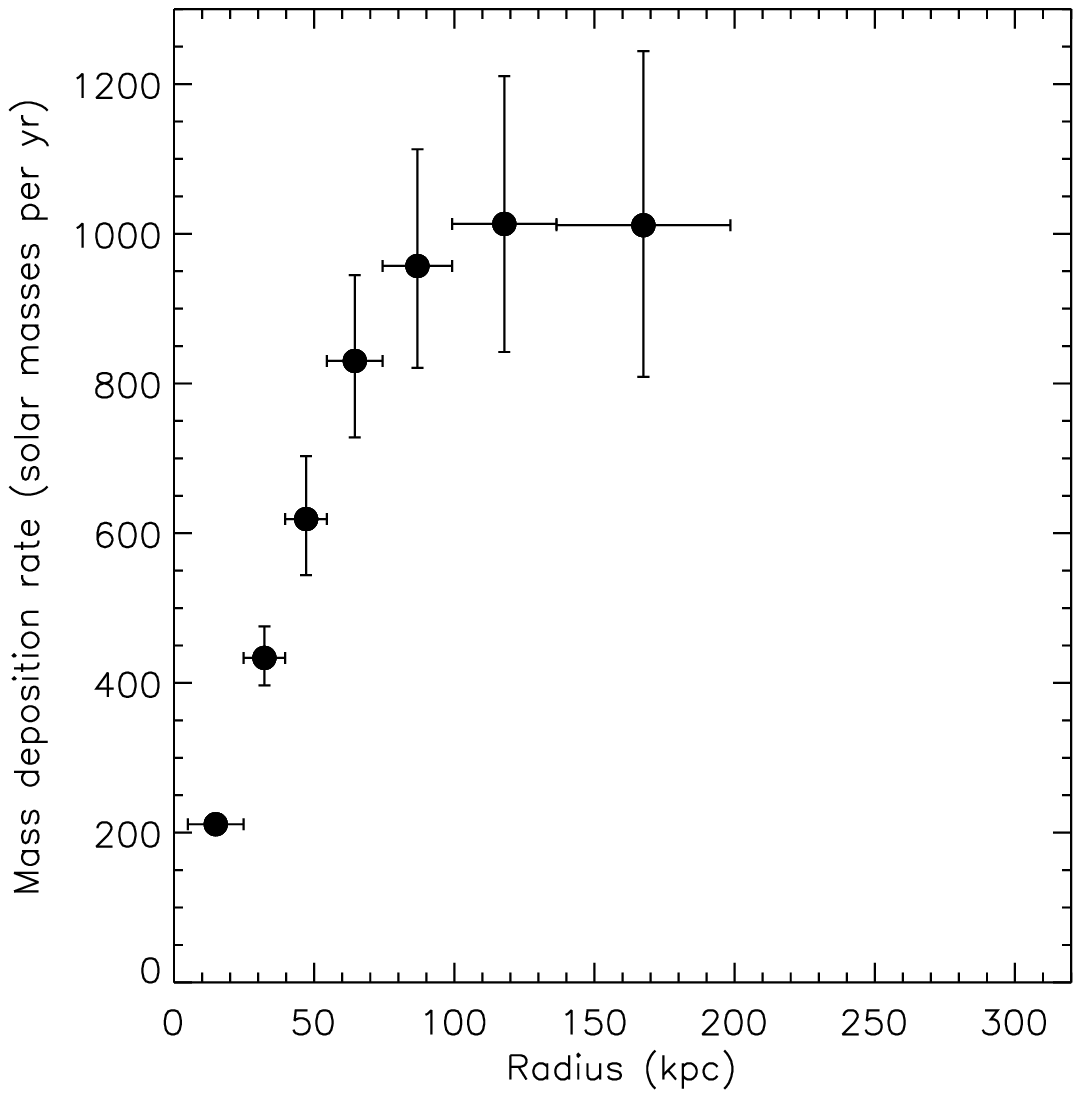}
\end{minipage}
\begin{minipage}[c]{0.32\linewidth}
\centering \includegraphics[width=\linewidth]{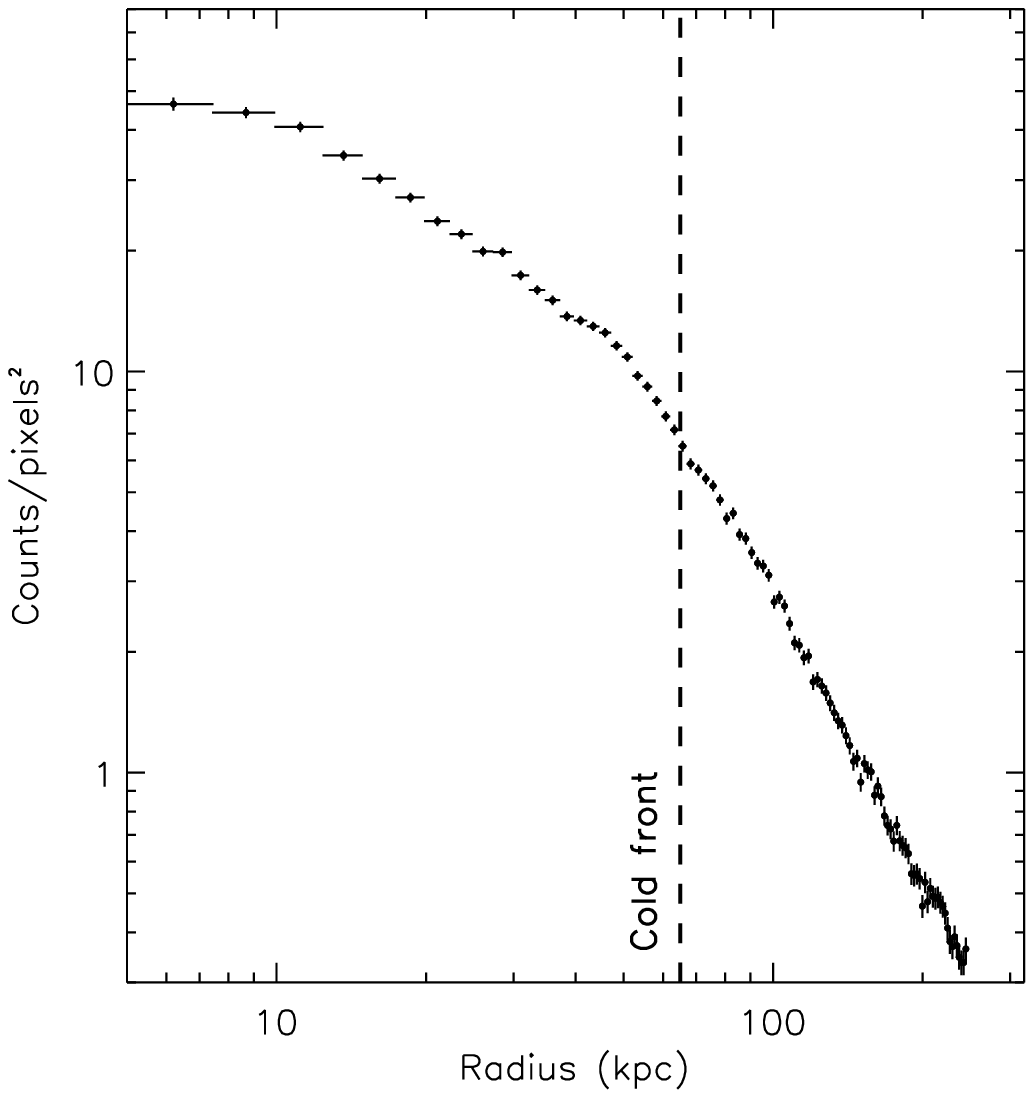}
\end{minipage}
\caption[]{Left - Cooling time profile derived from the \textit{Chandra} deprojected thermodynamic quantities. The cooling radius corresponds to the radius where the cooling time is equal to 7.7 Gyrs, approximately 115 kpc. Middle - Mass deposition \textit{Chandra} rate profile (${\rm\dot{M}}_I$) in the absence of heating. Note that the cooling time drops below 1 Gyr within 30 kpc, and that the mass deposition rate within the cooling radius is approximately 1000 ${\rm M_\odot}$ yr$^{-1}$. Right - X-ray surface brightness \textit{Chandra} profile in the 0.5-7.0 keV energy range, illustrating the cold front located at a radius of $\approx65$ kpc (see Section 7.1).
}
\label{fig2}
\end{figure*}

\subsection{Optical observations}
RX~J1532.9+3021 was observed with the Hubble Space Telescope (HST) as part of the Cluster Lensing And Supernova survey with Hubble \citep[CLASH][]{Pos201225}. The right panel of Fig. \ref{fig1} shows the F814W image taken with the Advanced Camera for Surveys (ACS) Wide Field Channel (WFC). It contains the redshifted H$\alpha$ emission line (H$\alpha\lambda6562$) and highlights well the complex filamentary structure of the BCG. We also show the F475W and F625W images taken with the same camera later in this work to further highlight the filaments. These contain respectively the [O\thinspace{\sc ii}]$\lambda3732$ and [O\thinspace{\sc iii}]$\lambda5007$ lines. 

We use the F225W image taken with the Wide Field Camera 3 (WFC3) UVIS imager to compute a UV star formation rate (SFR) associated with the galaxy. The extended filaments are not seen in this filter, which corresponds to a rest-wavelength of $\approx1737~\AA$. We derive the total SFR associated with the BCG within a circular region centred on the galaxy with a radius of 3.5$''$ ($\approx18$ kpc). Beyond this region, no significant UV emission is detected. We use the \citet{Kal2005440} value for the Galactic reddening assuming $N_{\rm H}=(2.21\pm0.09)\times10^{21}$ A$_{\rm V}$ \citep{Guv2010400} and R=3.1, to correct the HST F225W, F275W and F336W images for Galactic reddening. This results in an E(B-V) of 0.0335$\pm$0.0015 and we use the reddening curve of \citet{Car1989345}. 

To estimate the intrinsic reddening in the galaxy, we then make use of the F275W and F336W images taken with the same camera, while assuming that the monochromatic UV luminosity is flat in the $1500~$\AA $- 2800~$\AA wavelength range \citep{Ken1998498}. This allows us to derive an intrinsic flux density in the F225W band, and therefore a SFR for the galaxy in this band. We correct the flux density on a pixel-to-pixel basis. The average intrinsic reddening correction is E(B-V)=0.102, peaking at E(B-V)$\sim0.2$ in the central 0.5 arcsec. We note here that there are significant uncertainties arising both from the reddening corrections and the background determination of the UV filters. These dominate the uncertainties from counting statistics. We propagate them in our determination of the error for the SFR. We also assume that the UV emission is dominated by star formation, and not line emission. 

Taking these factors into account, the flux density corrected for Galactic reddening is $f_{\rm F225W}= (6.0\pm 2.9)\times10^{-28}$ ergs cm$^{-2}$ s$^{-1}$ Hz$^{-1}$. After correction for intrinsic reddening, the flux density is $f_{\rm F225W}= (1.2\pm0.6)\times10^{-27}$ ergs cm$^{-2}$ s$^{-1}$ Hz$^{-1}$ leading to a SFR of 76$\pm38~M_{\rm \odot}$ yr$^{-1}$ \citep{Ken1998498}.

\section{Large-scale structure imaging analysis}
RX~J1532.9+3021 is a highly X-ray luminous, and therefore massive cluster of galaxies. Based on the 2001 \textit{Chandra} observations, \citet{Man2010406} estimate that its total mass, within a radius where the mean enclosed density is 500 times the critical density at the cluster's redshift ($r_{\rm 500}$= 1.3 Mpc) is $(9.5\pm1.7)\times10^{14}M_{\rm \odot}$. 

RX~J1532.9+3021 is also a strong cool core cluster. We show this in Fig. \ref{fig2}, where the cooling time profile of the cluster is derived following the following equation:

\begin{equation}
t_{\rm cool} = \frac{5}{2}\frac{1.9~n_{\rm e}~kT~V}{L_{\rm X}} \,  .
\label{eq1}
\end{equation}

Here, $n_{\rm e}$ is the electron density, $kT$ is the gas temperature, $L_{\rm X}$ is the gas X-ray luminosity and $V$ is the gas volume contained within an annulus. Annuli were selected such that they each contained 8000 background-subtracted counts. We deprojected the spectra using the standard {\sc projct} mixing model in {\sc xspec}, which simultaneously fits the projected spectra from all annuli. For each annulus, we then fitted an absorbed {\sc mekal} model and let the temperature, abundance and normalization parameters free to vary while using C-statistics. Finally, we calculated the cluster luminosities within each of the annuli over the 0.01$\keV$ to 50$\keV$ energy range.
\begin{figure}
\begin{minipage}[c]{0.99\linewidth}
\centering \includegraphics[width=\linewidth]{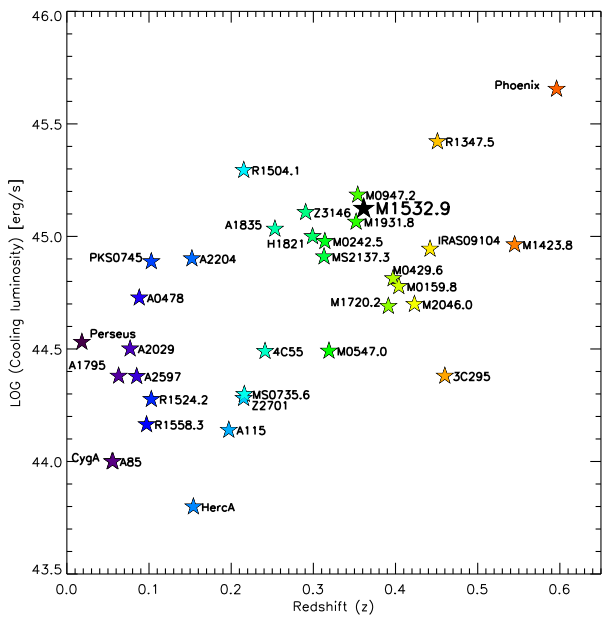}
\end{minipage}
\caption[]{Cooling luminosities in the $0.1-2.4$ keV energy range for a sample of strong cool core clusters. Note that, most harbour X-ray cavities and that RX~J1532.9+3021 lies among the strongest cool core clusters known (see black star). Colors go from purple to red with increasing redshift. Note that the error bars are included but are smaller than the symbols.}
\label{fig3}
\end{figure}

We define the cooling luminosity ($L_{\rm cool}$) as the $0.1-2.4\keV$ luminosity corrected for Galactic absorption within the cooling radius ($r_{\rm cool}$). We adopt the same definition for $r_{\rm cool}$ as \citet{Raf2006652}, which is taken as the radius where the cooling time is equal to the $z=1$ look-back time. We use the same definition so that we can directly compare our results with theirs. For our cosmology, this corresponds to $t_{\rm cool}=7.7$ Gyrs, and therefore $r_{\rm cool}\approx115$ kpc and $L_{\rm cool}=(13.7\pm0.1)\times10^{44}\ergps$. We also tested the Direct Spectral Deprojection (DSD) method of \citet{San2007381}, obtaining results consistent with those obtained using the {\sc projct} mixing model. 

We furthermore compute the mass deposition rate profile (${\rm\dot{M}}_I$) in the absence of heating using the method of \citet{Whi1997292} which includes the work done by gravity, and using our deprojected thermodynamic quantities. The results are shown in Fig. \ref{fig2}. The mass deposition rate within $r_{\rm cool}$ is around 1000${\rm M_\odot}$ yr$^{-1}$. Such a high rate would result in a starburst, which is clearly not seen. Fig. \ref{fig2} also shows the X-ray surface brightness profile of the cluster in the $0.5-7$ keV energy range. Each point represents the average surface brightness within a given annulus. There is a clear break in the profile at $\approx65$ kpc, which is characteristic of a cold front. We further analyse this feature in Sections 5.2 and 7.1.

In Fig. \ref{fig3}, we compare RX~J1532.9+3021 to other strong cool core clusters. We plot the cooling luminosity as a function of redshift for the sample of \citet{Hla2013431}, which derives the cooling luminosities in an identical way for known massive galaxy clusters with X-ray cavities. We also add RXC J1504.1-0248 \citep{Ogr2010406}, RXC J1347.5-1145 and the Phoenix cluster \citep{McD2012Nat}. Fig. \ref{fig3} shows that RX~J1532.9+3021 is one of the strongest cool core clusters known, with a cooling luminosity 4 times larger than the Perseus cluster.

\begin{figure*}
\centering
\begin{minipage}[c]{0.99\linewidth}
\centering \includegraphics[width=\linewidth]{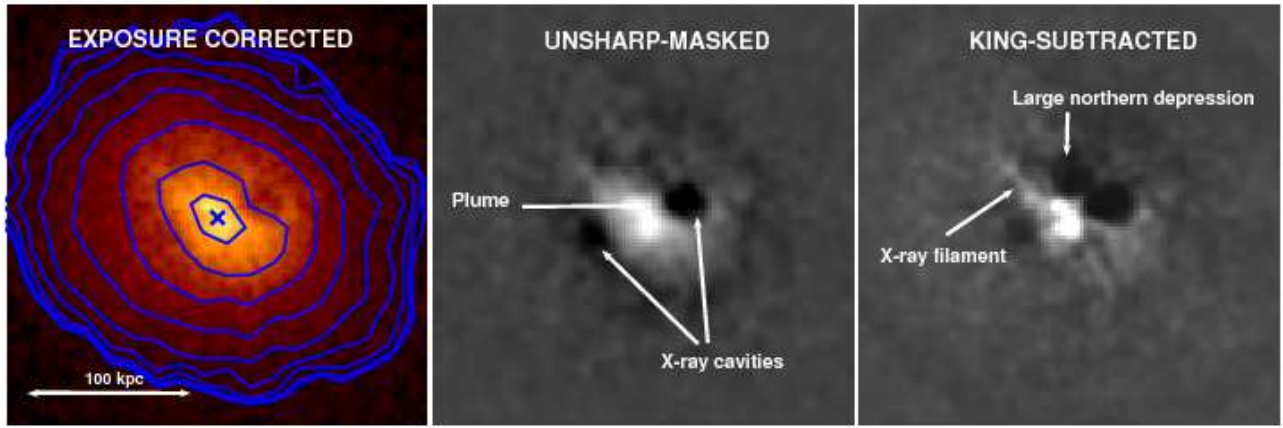}
\end{minipage}
\caption[]{Right - Exposure-corrected $0.5-7\keV$ \textit{Chandra} image, same as in Fig. \ref{fig1}, but zoomed in and with logarithmic surface brightness contours in blue. Middle - Unsharp-masked \textit{Chandra} image where a 2D gaussian smoothed image ($\sigma=10$ pixels) was subtracted from a less smoothed image ($\sigma=1.5$ pixels). Right - Smoothed image (2D gaussian with $\sigma=1.3$ pixels), subtracted by a King model of the cluster emission (see Section 4). The centre is chosen as the location of the optical nucleus, shown with the blue cross. Interesting features are highlighted.  }
\label{fig4}
\end{figure*}

\begin{table}
\caption[]{X-ray derived properties of RX~J1532.9+3021. ${\rm\dot{M}}_S$ measures the mass deposition rate as observed from the X-ray spectra, and ${\rm\dot{M}}_I$ measures of the mass deposition rate in the absence of heating. $^a$\citet{Man2010406}.}
\resizebox{8.75cm}{!} {
\begin{tabular}{llclr}
\hline
& Region & Value & Units & Telescope \\
\hline
\hline
${\rm M}_{\rm cluster}$ &${\rm r<r_{\rm 500}}$ & $9.5\pm1.7$ & $10^{14}M_{\rm \odot}$ & $Chandra^a$\\


${\rm L}_{\rm cool}$ &${\rm r<r_{\rm cool}}$ & $13.7\pm0.1$ & $10^{44}\ergps$ &\textit{Chandra}\\

${\rm\dot{M}}_I$ &${\rm r<r_{\rm cool}}$ & $\approx$1000  & ${\rm M_\odot}$ yr$^{-1}$&\textit{Chandra}\\

${\rm\dot{M}}_I$ &${\rm r< 50~{\rm kpc}}$ & $\approx$600  & ${\rm M_\odot}$ yr$^{-1}$&\textit{Chandra}\\

${\rm\dot{M}}_S$ &${\rm r<r_{\rm cool}}$ & $251\pm{60}$ & ${\rm M_\odot}$ yr$^{-1}$& \textit{Chandra} \\

${\rm\dot{M}}_S$ &${\rm r<50~{\rm kpc}}$ & $192\pm43$ & ${\rm M_\odot}$ yr$^{-1}$& \textit{Chandra}\\

${\rm\dot{M}}_S$ &... & $<305$ & $ {\rm M_\odot}$ yr$^{-1}$& RGS \textit{XMM}\\

\hline
\end{tabular}}
\vspace{0.3cm}
\label{tab3}
\end{table}

Finally, we searched for evidence of cooling gas within the central 50 kpc, as well as within $r_{\rm cool}$. We denote this measurement as ${\rm\dot{M}}_S$, a quantity that measures the mass deposition rate as observed from the X-ray spectra. We use a two component model consisting of a {\sc mekal} and {\sc mkcflow} component, the latter of which is appropriate for gas cooling at constant pressure from an upper temperature down to a lower temperature. We tie the upper temperature and metallicity of the {\sc mckflow} component to those of the {\sc mekal} component and to help constrain the fit, we further set the low temperature to 0.1 keV. The results, computed for both the \textit{Chandra} and the RGS \textit{XMM} data are shown in Table \ref{tab3}. For the RGS data, we fitted the first-order data between 7 and 26 \AA\ , and for the \textit{Chandra} data we used the $0.5-7$ keV energy range. We were only able to derive an upper limit for the cooling rate from the RGS \textit{XMM} observations due to the limited quality of the data.

We also investigated the possibility of multi-temperature gas existing in the central 50 kpc, as well as within $r_{\rm cool}$, by fitting an absorbed two-temperature model to the \textit{Chandra} data with the abundances of both models tied together. We find clear evidence for a cooler component present in the spectrum. For the region within 50 kpc, the cooler (kT=$0.54\pm0.04$ keV) component contributes $\approx2$ per cent of the flux compared to the hotter ($4.59\pm0.09$ keV) component, whereas for the region within $r_{\rm cool}$, the cooler (kT=$0.5\pm0.1$ keV) component contributes $\approx1$ per cent of the flux compared to the hotter ($5.41\pm0.08$ keV) component. This 0.5 keV low-temperature floor has been seen in many other groups and clusters \citep[e.g.][]{McN2000534}. \citet{Wer2013767} recently suggested that the lack of very soft X-ray emission could be due to absorption by the cold gas in the optical line emitting filaments if the integrated hydrogen column density is high enough.

\begin{figure*}
\centering
\begin{minipage}[c]{0.35\linewidth}
\centering \includegraphics[width=\linewidth]{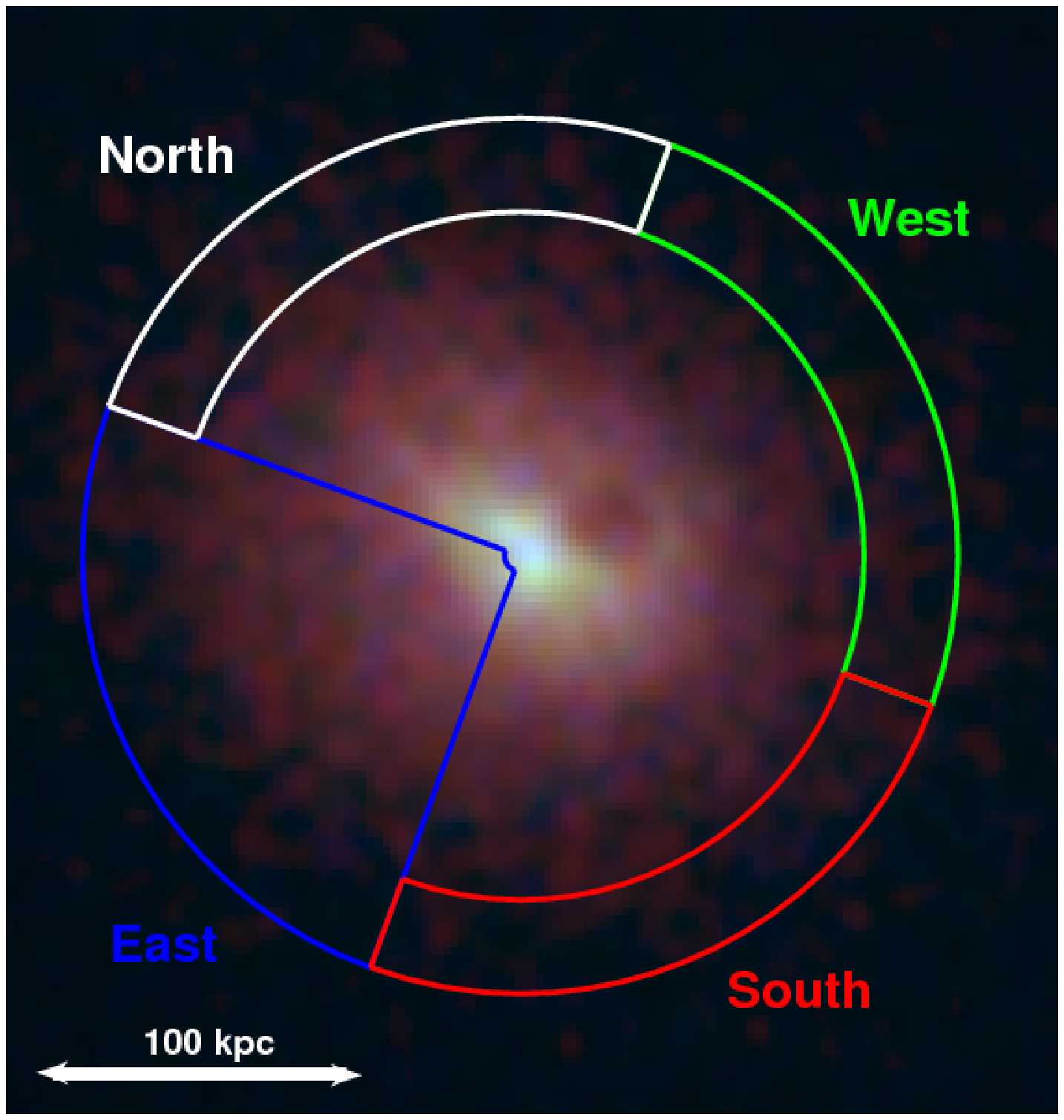}
\end{minipage}
\begin{minipage}[c]{0.55\linewidth}
\centering \includegraphics[width=\linewidth]{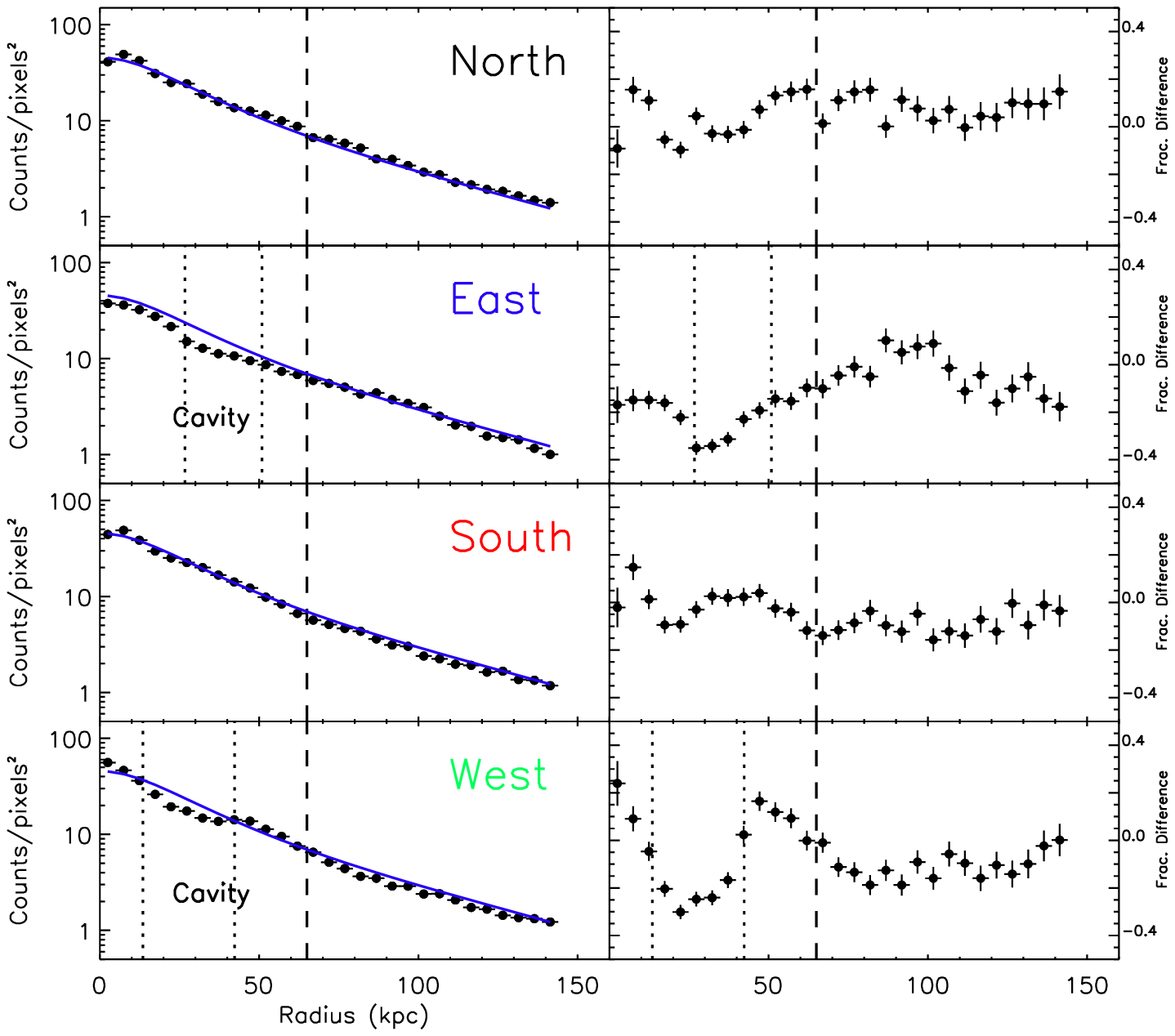}
\end{minipage}
\caption[]{Left - Three color image of RX~J1532.9+3021: $0.3-0.8\keV$ (red), $0.8-3.0\keV$ (green) and $3.0-8.0\keV$ (blue), with four sectors overplotted. Right - Surface brightness profile along each sector. The blue line is the best-fitting King plus constant model of the combined north plus south sector (those less affected by the X-ray cavities). Also shown, are the locations of each cavity (between the dotted lines) and the located of the cold front (dashed line).}
\label{fig5}
\end{figure*}

\begin{figure*}
\centering
\begin{minipage}[c]{0.7\linewidth}
\centering \includegraphics[width=\linewidth]{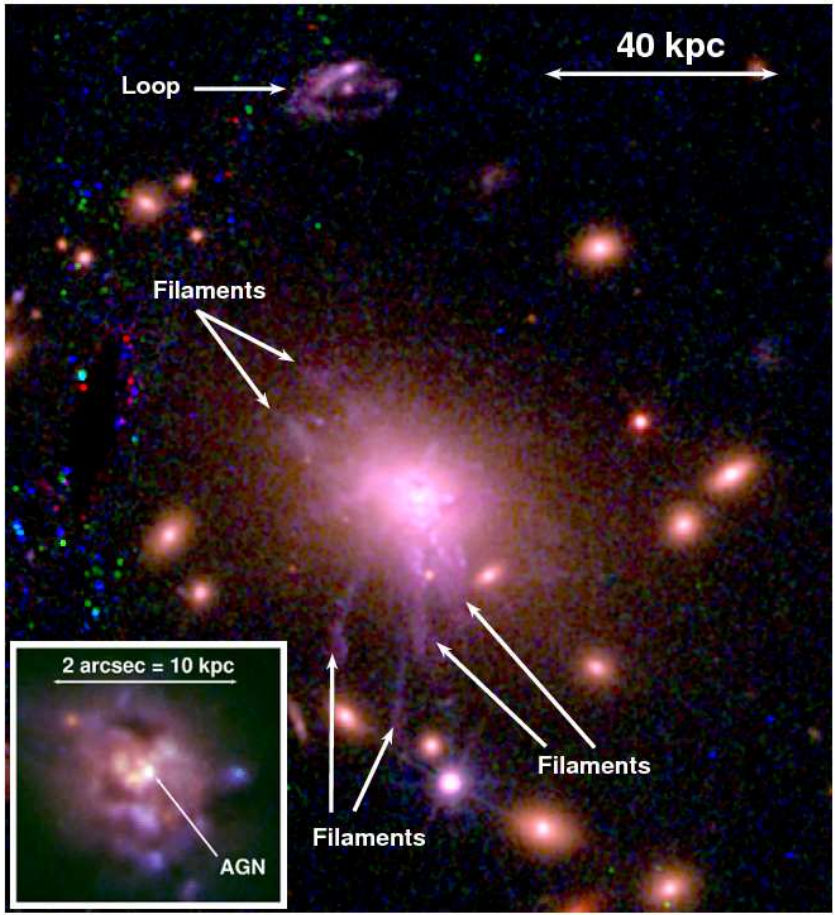}
\end{minipage}
\caption[]{Combined HST F814W (red), F625W (green) and F475W (blue) images of the central regions of RX~J1532.9+3021. The inset to the bottom-left focuses on the optical nucleus. }
\label{fig6}
\end{figure*}

\section{Imaging analysis of the core}

\begin{table*}
\caption[]{X-ray cavity properties in RX~J1532.9+3021.}
\resizebox{18cm}{!} {
\begin{tabular}{lccccccccc}
\hline
\hline
Cavity & $R_{\rm l}$ & $R_{\rm w}$ & $R_{\rm dist}$ & $t_{\rm sound}$ & $t_{\rm buoy}$ & $t_{\rm refill}$  & $P_{\rm sound}$ & $P_{\rm buoy}$ & $P_{\rm refill}$ \\
 & (kpc) & (kpc) & (kpc) & ($10^7$ yr) & ($10^7$ yr) & ($10^7$ yr) & ($10^{44}\ergps$) & ($10^{44}\ergps$) & ($10^{44}\ergps$) \\
\hline
West & 14.4$\pm$2.9 & 17.3$\pm$3.5 & 28 & 2.8$\pm$0.7 & 3.1$\pm$0.5 & 6.3$\pm$0.7 & 19.4$\pm$10.0 & 17.4$\pm$8.3 & 8.5$\pm$3.9 \\
& & & & & & & & & \\
East & 12.1$\pm$2.4& 14.9$\pm3.0$ & 39 & 3.4$\pm$0.9 & 6.6$\pm$0.9 & 8.2$\pm$0.7 & 9.4$\pm$4.9 & 4.8$\pm$2.2 & 3.9$\pm$1.8\\
\hline
\end{tabular}}
\label{tab4}
\end{table*}

Fig. \ref{fig1} clearly reveals the presence of an X-ray cavity to the west, as well as hints of a second X-ray cavity to the east. The second cavity becomes clear when we plot the unsharp-masked image (Fig. \ref{fig4}), where a 2D gaussian smoothed image ($\sigma=10$ pixels) was subtracted from a less smoothed image ($\sigma=1.5$ pixels). In Fig. \ref{fig5} we plot the surface brightness profiles along four different directions, as well as the fractional differences between the observed profile and a best-fitting King plus constant model per sector defined as 
\begin{equation}
I(r) = I_{\rm 0} \left(1 + \left(\frac{r}{r_{\rm 0}}\right)^2\right)^{-\beta} + C_{\rm 0} \, 
\label{eq9}
\end{equation}

and derived only using the north plus south sector which are less affected by the cavities. Here, a 0 means no difference, whereas a 1 means a difference of 100 per cent. Fig. \ref{fig5} reveals decrements for each of the cavities (about 30 per cent, see also Table \ref{tab4}). The unsharp-masked image of Fig. \ref{fig4} also reveals a plume-like structure at the centre and we analyze this feature further in Section 5.1.

\begin{figure}
\centering
\begin{minipage}[c]{0.9\linewidth}
\centering \includegraphics[width=\linewidth]{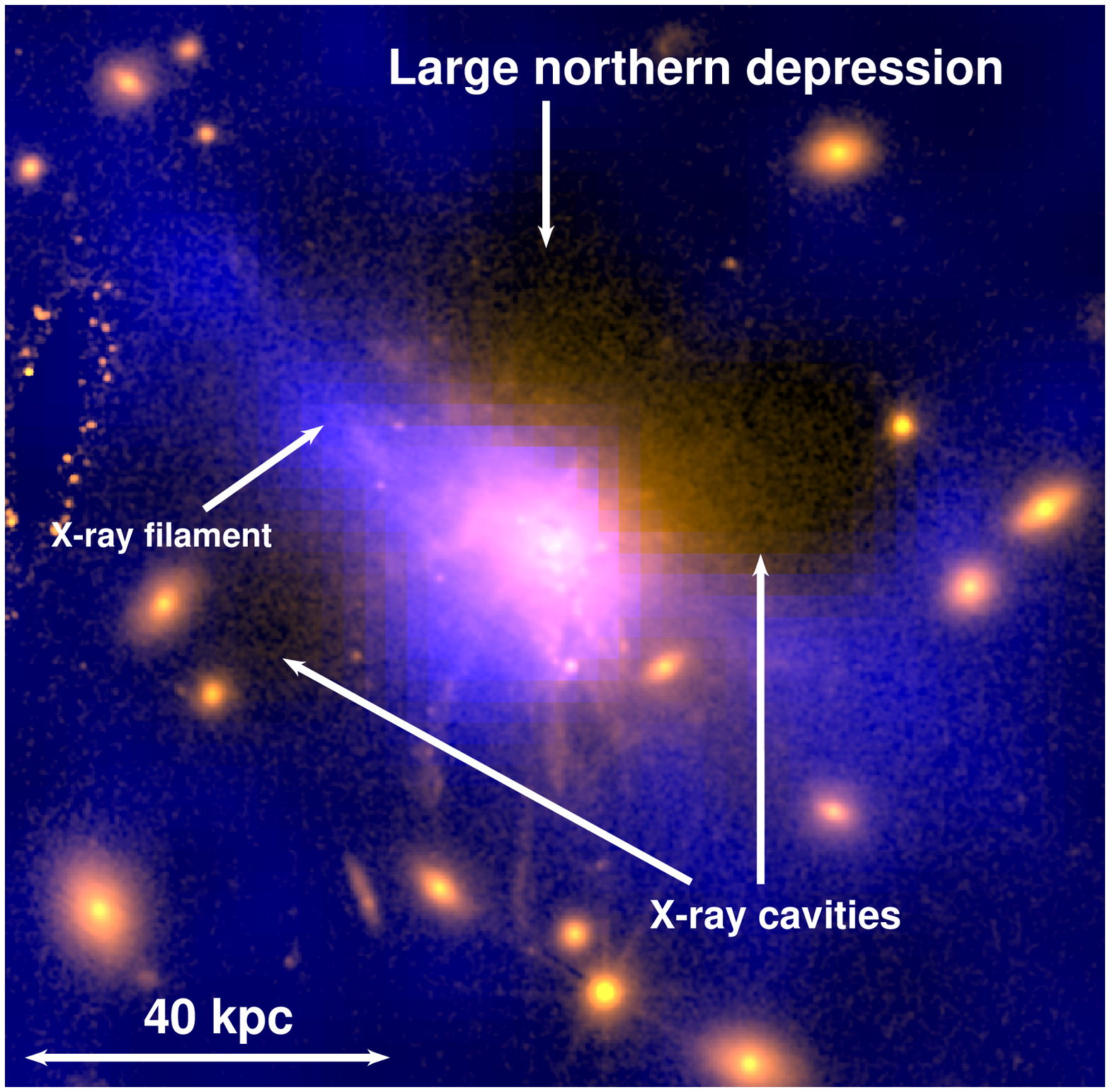}
\end{minipage}
\caption[]{Shown is the total optical HST image (yellow; resulting from the combination of the F435W, F475W, F606W, F625W, F775W, F814W and F850lp images), combined with the King-subtracted $0.5-7$ keV X-ray image of Fig. \ref{fig4} (blue). }
\label{fig6b}
\end{figure}

To further enhance the structures within the core, we developed a technique that consists of initially smoothing the \textit{Chandra} image with a 2D gaussian smoothed image ($\sigma=1.3$ pixels) and then dividing the image into 12 sectors centred on the nucleus. We choose such a large number of sectors to trace the azimuthal differences throughout the cluster. To each of these sectors, we fit a King model (Eq. \ref{eq9}) to the surface brightness profile and then replace the value in each pixel with the average of the King profile values of the two neighbouring sectors located at the same radius. We refer to this new image as the ``averaged" image. Finally, we subtract this averaged image from the original image, and show the results in the right panel of Fig. \ref{fig4}. The centre is chosen as the location of the nucleus, taken as the brightest point within the central 20 kpc of the \textit{HST} images, see the point source labelled as AGN in the inset of Fig. \ref{fig6}. Although the centre also coincides with the centre of the plume-like structure, our results remain the same even if we shift the centre to $\pm3$ pixels. We also obtain consistent results when we vary the number of sectors, as well as when we exclude the central 5 pixels in radius containing the plume-like feature. Fig. \ref{fig4} essentially reveals hints of a large northern depression. It may represent an older X-ray cavity or a region where they have accumulated, similar to what is observed in Perseus \citep{Fab2011418}. It could also result from leakage of relativistic particles from the western X-ray cavity. We further discuss these possibilities in Section 7.2.

Fig. \ref{fig6} shows the combined HST F814W (red), F625W (green) and F475W (blue) images. The BCG is surrounded by a complex array of filaments, similar to what is observed in the Perseus cluster. The most prominent filaments are located in the southern and north-eastern directions. These extend to some 50 kpc in size, making them some of the largest filamentary complexes known around BCGs. The cavities we have identified are located to the east and west, and small filaments trailing after them, and in particular after the western cavity, can be seen through the HST images. This is consistent with the idea that some filaments originate from gas being dragged out of the central galaxy by the AGN outflows \citep[e.g.][]{Chu2001554}. As for the southern filaments, their existence might indicate the presence of an older outflow which has risen buoyantly and can no longer be clearly seen with the current X-ray data. We analyse this further in Section 7.2.

Interestingly, the King-subtracted X-ray image shown in Fig. \ref{fig4} reveals the presence of a faint X-ray filament in the north-eastern direction. To highlight this better, we show the combined King-subtracted X-ray and \textit{HST} images in Fig. \ref{fig6b}, and point to the location of the X-ray filament. Hints of the X-ray filament can also be seen in the soft $0.3-1$ keV image, without applying any techniques to enhance the deviations.

Finally, we mention the presence of a bright region to the north of the BCG, identified in Fig. \ref{fig6} as the ``loop''. The region is located $\approx$80 kpc (projected) from the central AGN, and consists of two bright knots, surrounded by a diffuse loop-like structure. It is also particularly blue, suggesting strong star formation. It could be the remnant of a galaxy-galaxy merger located within the cluster, a lensed star-forming galaxy located behind the cluster, or, due to the similar colors, a distant filament that has been dragged out by a past AGN outburst. To test this, we constructed spectral energy distributions (SEDs) for its northern knot, as well as for its diffuse loop-like part, and for several of the filaments surrounding the BCG. The SED of the loop is very different than those of the filaments. It peaks in the UV, whereas those of the filaments peak in the F814W filter which contains the redshifted H$\alpha$ emission line. The SED of the knot is also particularly flat, consistent with it being an AGN. The structure is therefore most likely a galaxy and not a distant filament. There are however no redshifts available on the source, making it difficult to say whether it is a galaxy that belongs to the cluster, or whether it is simply a lensed background galaxy.

\section{X-ray spectral analysis}

\subsection{Thermodynamic maps}
To study the temperature and metallicity distribution across the cluster, we bin different regions together using a contour binning algorithm which follows the surface brightness variations \citep[][]{San2006371CB}. 

\begin{figure}
\centering
\begin{minipage}[c]{0.8\linewidth}
\centering \includegraphics[width=\linewidth]{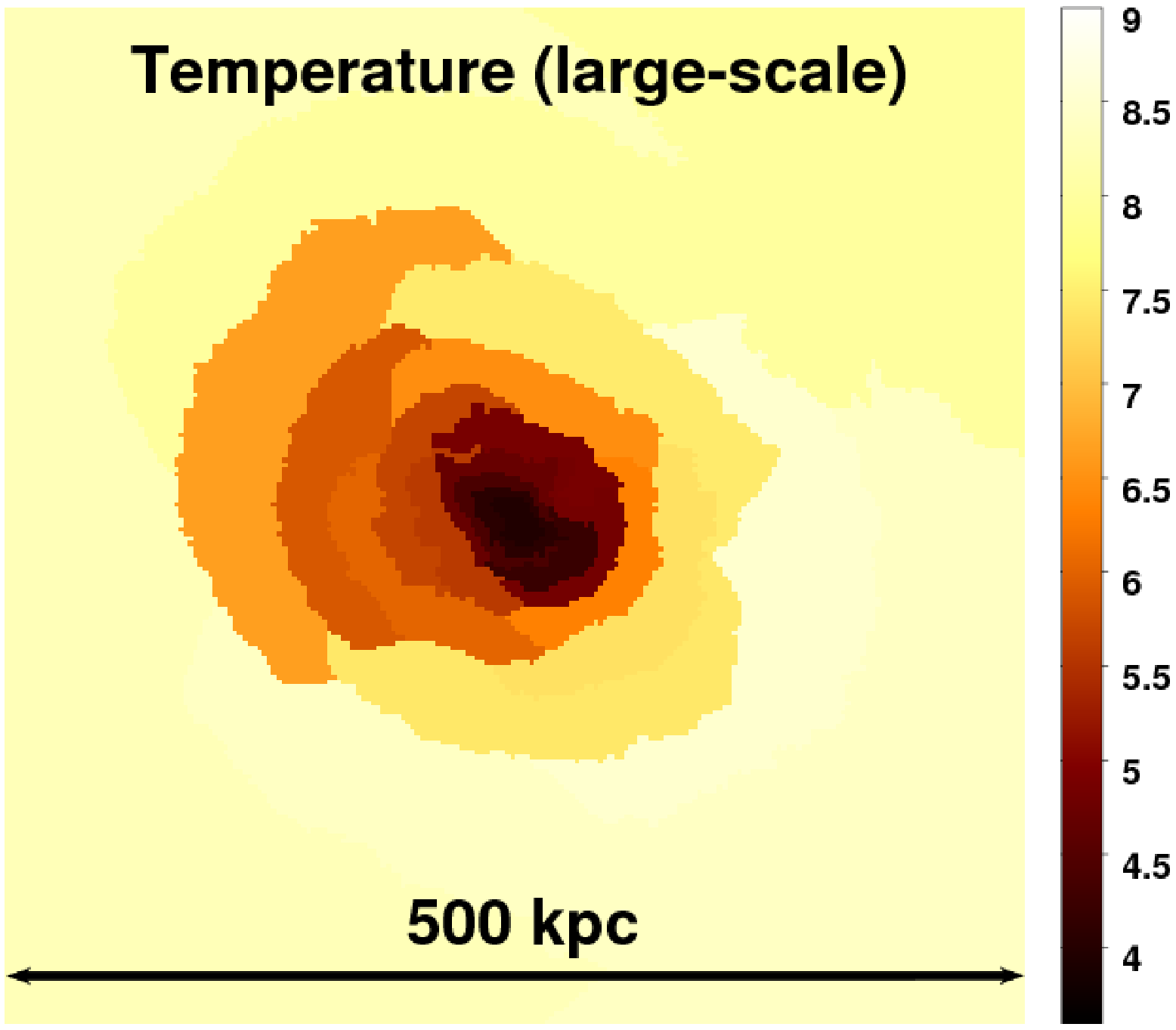}
\end{minipage}
\begin{minipage}[c]{0.8\linewidth}
\centering \includegraphics[width=\linewidth]{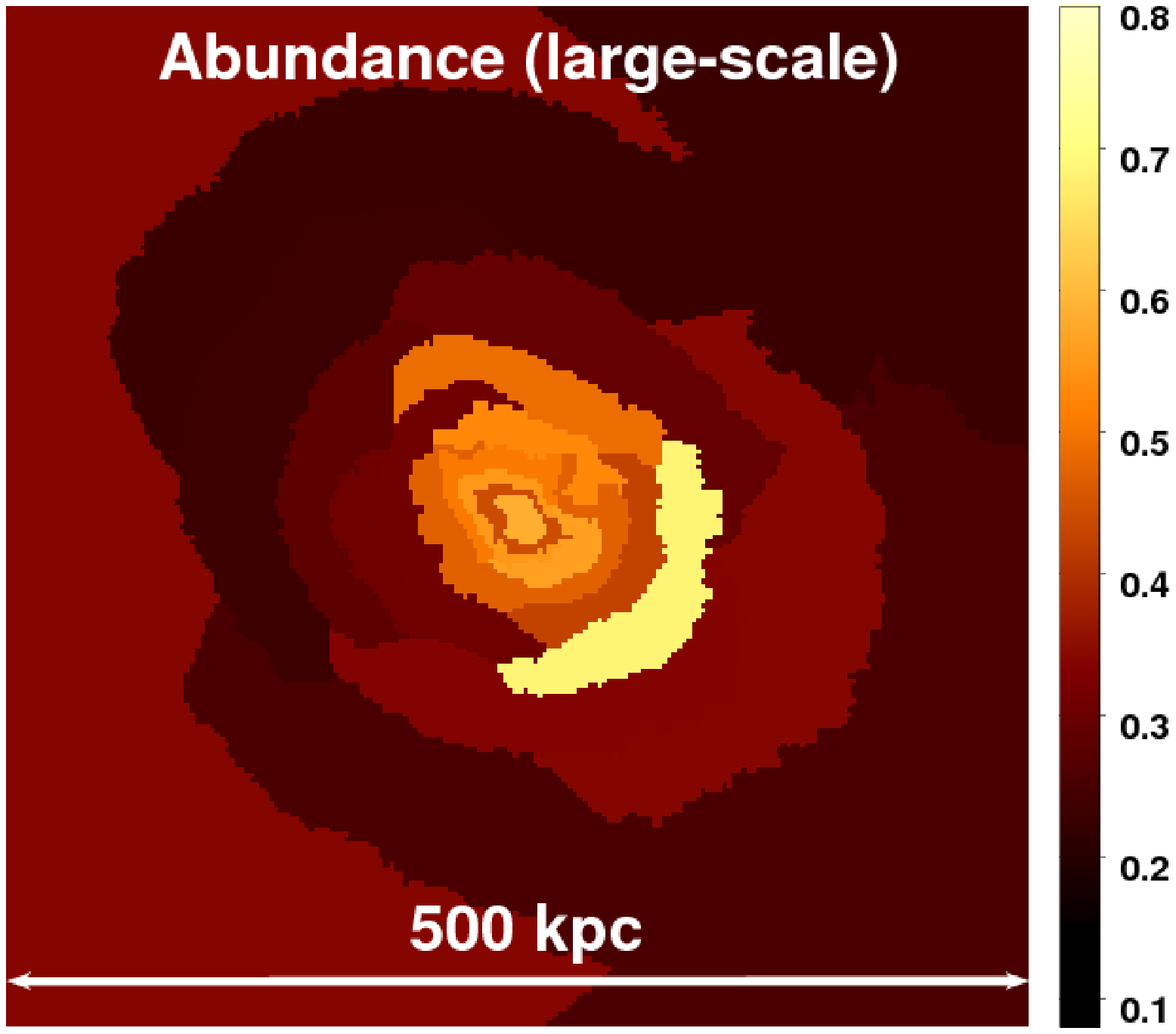}
\end{minipage}
\caption[]{Large-scale temperature and abundance map, obtained from binning the regions with a minimum signal-to-noise ratio of 60 (3600 counts). Note the temperature difference between the east and west side.}
\vspace{0.05in}
\label{fig8}
\end{figure}

We begin by analysing the large-scale structures within the central $500\times500$ kpc. We bin the regions so that they each contain at least 3600 counts, initially smooth the image with a signal-to-noise ratio of 15 and restrict the lengths to be at most 2 times that of a circle with the same area. We extract a $0.5-7\keV$ spectrum for each region and fit an absorbed single {\sc mekal} model to the data while using C-statistics. We left the temperature, abundance and normalisation parameters free to vary. The resulting temperature and metallicity maps are shown in Fig. \ref{fig8}. For the temperature, the errors on each value vary from $4$ per cent in the inner regions to $8$ per cent in the outer regions, and for the abundance they vary from $16$ per cent (inner) to $36$ per cent (outer). The average temperature in the eastern direction (over 4 bins) is $6.0\pm0.2$ keV, whereas the temperature to the west, located within bins of similar radius, is $7.4\pm0.2$ keV. The east direction is therefore statistically cooler than the west, and may be the result of a past merger between the cluster and a cooler subcluster along the east-west direction. There is also a south-west region that appears more metal rich (0.7$Z_{\rm \odot}$) than the average value seen throughout the cluster (0.4$Z_{\rm \odot}$), similar to what is seen in the Ophiuchus cluster \citep{Mil2010405}, but this increase lies within $2\sigma$ of the neighbouring bins. Also seen, is a contrast in the temperature structure to the west, where the temperature jumps from 5$\keV$ to 7$\keV$ at a radius of around 65 kpc. This could indicate the presence of a cold front. We further investigate this in the following section. Note that the evidence of multi-phase gas found in Section 3 mostly contributes to the inner 50 kpc, and does not affect the large-scale temperature and abundance map significantly beyond the inner few bins. 

\begin{figure}
\centering
\hspace{-0.2cm}
\begin{minipage}[c]{0.59\linewidth}
\centering \includegraphics[width=\linewidth]{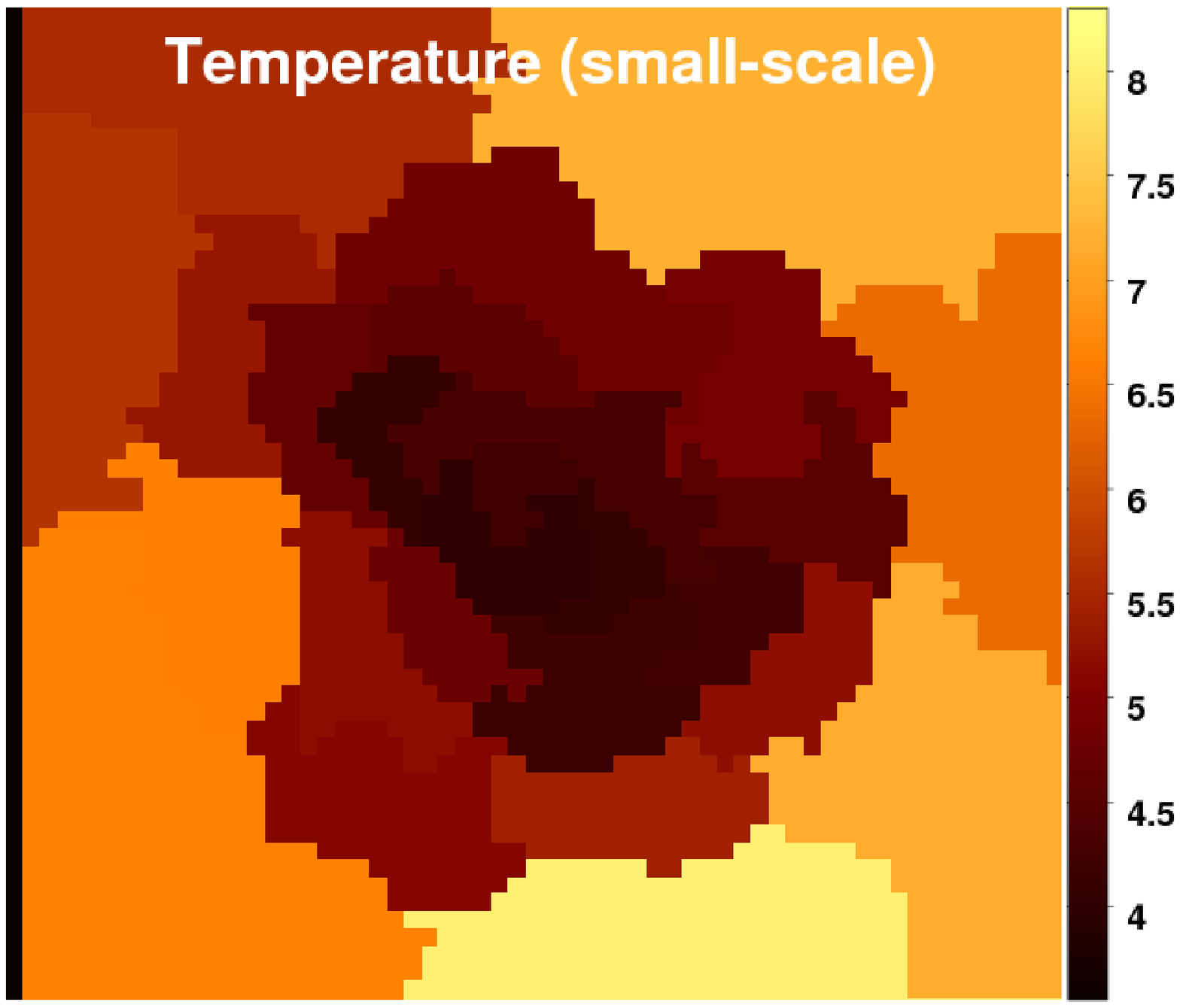}
\end{minipage}
\hspace{-0.8cm}
\begin{minipage}[c]{0.47\linewidth}
\centering \includegraphics[width=\linewidth]{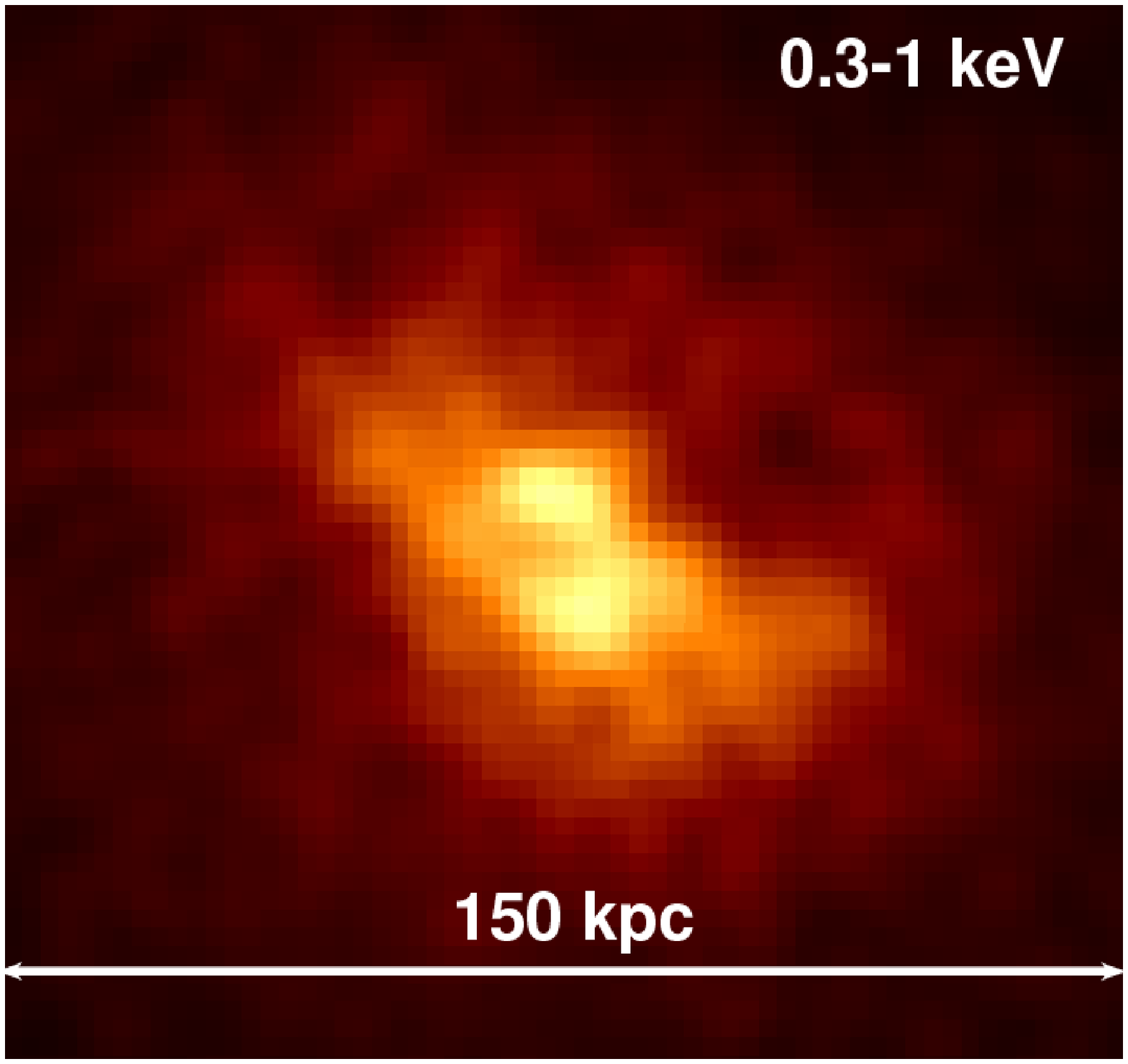}
\end{minipage}
\vspace{0.05in}
\caption[]{Small-scale temperature map (left), obtained from binning the regions along the surface brightness contours of the soft $0.3-1$ keV X-ray image (right), and then extracting the spectrum from the full $0.5-7$ keV energy band. The resulting regions have a signal-to-noise ratio of approximately 30 (900 counts), and clearly reveal the cool core.}
\label{fig9}
\end{figure}

Next, we analyze the inner $150\times150$ kpc region which shows interesting features especially at soft X-rays (see right panel of Fig. \ref{fig9}). To highlight the thermodynamic properties of these features, we bin the regions so that they follow the surface brightness contours of the soft $0.3-1.0\keV$ image. We initially smooth the image with a gaussian function ($\sigma=1.3$ pixel), and then bin the regions so that they each contain at least 400 counts. We also restrict the lengths to be at most 1.3 times that of a circle with the same area. However, we extract the spectrum from each region in the full $0.5-7\keV$ range, which will contain more than 400 counts (around 900 counts). We use C-statistics and show the results in Fig. \ref{fig9}. Here, the errors vary from $7$ per cent in the inner regions to $11$ per cent in the outer regions for the temperature, and from $30$ per cent (inner) to $55$ per cent (outer) for the abundance. Due to the large error bars on the abundance, we only focus on the temperature structure. The number of counts from the current data are also not sufficient to fit a two-temperature model to each of the regions selected. We would need at least two to three times the number counts per region to provide a good fit (or regions two to three times larger). Therefore, although we found evidence for multi-temperature gas within the inner 50 kpc, we focus here on the small-scale temperature structure, choose small regions containing around 900 counts each, and only fit a single {\sc mekal} model to each region. Both the small-scale and large-scale temperature maps show that the cluster has a cool core, with the temperature dropping from 8$\keV$ in the outskirts to 4$\keV$ in the central 100 kpc. 

The plume-like feature mentioned in Section 4 stands out clearly in the right panel of Fig. \ref{fig9}, indicating that the plume most likely contains cooler material. However, when fitting an absorbed {\sc mekal} model to the \textit{Chandra} spectrum of the plume, we find that the temperature and abundance are consistent to a 1$\sigma$ level with those of a surrounding region, located within the same radius as the plume but excluding the plume. The data therefore remain too poor to determine clearly how much cooler the plume is, and deeper observations are needed. 

\begin{figure}
\begin{minipage}[c]{1.2\linewidth}
\hspace{-0.6in}
\includegraphics[width=\linewidth]{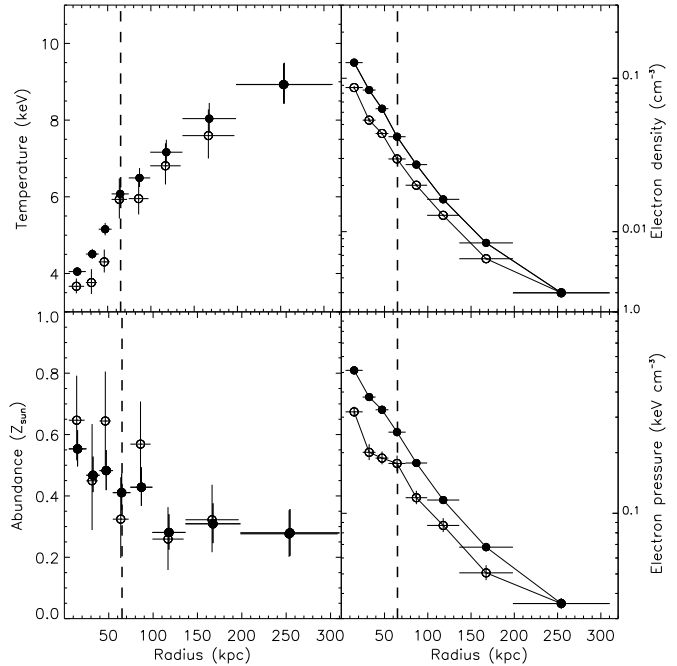}
\end{minipage}
\vspace{-0.2in}
\caption[]{Projected (filled symbols) and deprojected (empty symbols) temperature, metallicity, electron density and electron pressure profiles, as derived by selecting annuli with a minimum signal-to-noise ratio of 90 ($8000$ counts). The dashed line represents the location of the cold front.   }
\label{fig11}
\end{figure}

\begin{figure*}
\centering
\begin{minipage}[c]{0.85\linewidth}
\centering \includegraphics[width=\linewidth]{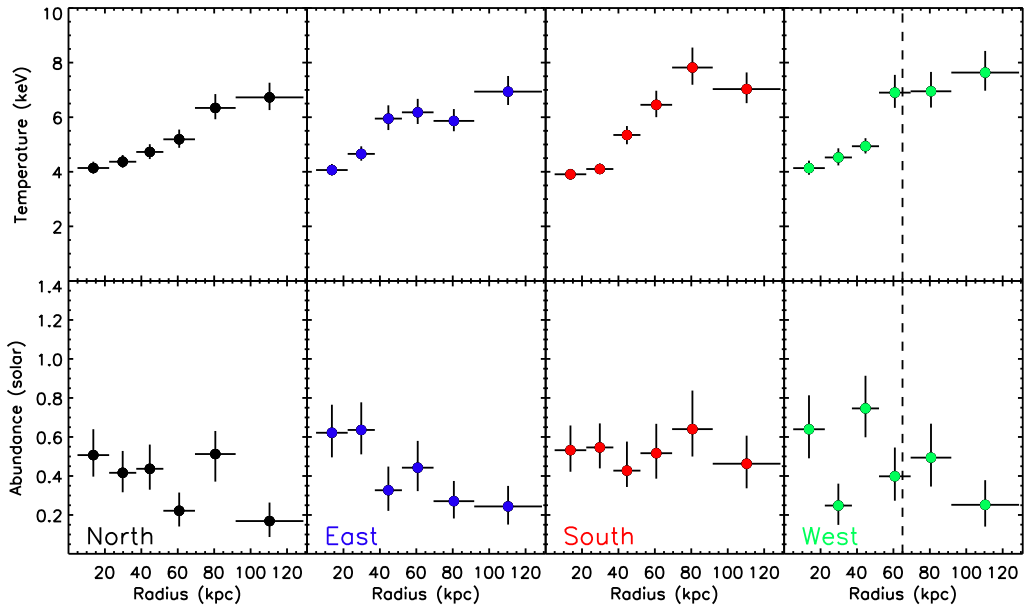}
\end{minipage}
\caption[]{Projected temperature, metallicity, electron density and electron pressure profiles along the four sectors shown in Fig. \ref{fig5}. Each annulus contains a minimum signal-to-noise ratio of 45 ($\approx2000$ counts). The dashed line highlights the cold front (see Section 7.1).  }
\label{fig10}
\end{figure*}

\subsection{Thermodynamic profiles}
We now proceed to do a more in depth analysis of the thermodynamic properties of the cluster, by computing thermodynamic profiles.

We first select annular regions containing roughly a signal-to-noise ratio of 90 ($8000$ counts), and push the analysis to larger radii. The projected and deprojected temperature, metallicity, electron density and electron pressure profiles are shown in Fig. \ref{fig11}. For each spectrum, we fit an absorbed {\sc mekal} model while using C-statistics and we deprojected the spectra using the standard {\sc projct} mixing model in {\sc xspec}. Note that the multi-temperature gas discussed in Section 3 only contributes to the inner 3 annuli, and the thermodynamic profiles of the hotter component are consistent with those shown in Fig. \ref{fig11}.

Fig. \ref{fig11} shows that beyond a radius of $100$ kpc, the projected metallicity profile decreases with radius from $Z\approx0.6$ solar to $Z\approx0.3$ solar. This figure also shows the characteristic drop in temperature with decreasing radius seen in cool core clusters. However, there is a distinct, interesting feature at a radius of 65 kpc (see dashed line). Here, the deprojected temperature increases from 4.5 keV to 6 keV, while the deprojected electron pressure profile breaks at this radius, such that the pressure remains constant across the front. The latter is a typical characteristic of cold fronts.

To further investigate the potential cold front, we compute thermodynamic profiles along the same four sectors as Fig. \ref{fig5}. We select regions so that they each contain a signal-to-noise ratio of $45$ ($2000$ counts), and do not correct for deprojection effects due to the limited number of counts. The number of counts are also not sufficient to fit a two-temperature model to the data. The results are shown in Fig. \ref{fig10}, and the strongest temperature jump observed lies along the western direction and a radius of 65 kpc. Interestingly, the front coincides with the edge of the western X-ray cavity as well as the edge of the radio mini-halo. We further discuss this front in Section 7.1.

\section{The central AGN}
\subsection{Radiative properties of the central AGN}

Using the \textit{Chandra} X-ray images, we searched for evidence of a central point source. The hard $2-8\keV$ image reveals three bright pixels in the central region that coincide with the nucleus as seen from the \textit{HST} image (RA=15$^h$32$^m$53.77$^s$, DEC=+30$^o$20$'$59.44$''$; see inset of Fig. \ref{fig6}). Here, the nucleus in the \textit{HST} image corresponds to the brightest point within the central 20 kpc. Taking a circular region with a radius of $1''$ centred on these bright pixels, we compare the total counts to the $3\sigma$ upper limit of the surrounding background. The background is taken as a surrounding annulus with an interior radius of 1$''$ and an outer radius of 2$''$, and separately as a surrounding annulus with an interior radius of 2$''$ and an outer radius of 4$''$. In both cases, the counts within the central $1''$ lie well within the $3\sigma$ upper limit of the background, and there is no clear evidence for an excess associated with a central point source.

To further show this, we calculate the hardness ratio, defined as (H-S)/(H+S). Here, H represents the hard number of counts within a certain region as seen from the $2-8\keV$ image, and S represents the soft number of counts from the same region, but taken from the $0.5-2\keV$ image. We compute the hardness ratio for the central $1''$ region and find a ratio of $-0.50\pm0.04$, where the uncertainty is determined by assuming that the counts are governed by poisson statistics. The hardness ratio of the surrounding $2-4''$ region is $-0.57\pm0.02$. As seen, the ratios lie within $2\sigma$ of each other, and there is no clear evidence that the central 1$''$ has a harder spectrum than the surrounding cluster emission.

The spectrum of the central 1$''$ region also shows no clear evidence for the presence of non-thermal emission. By taking this region and a surrounding annulus with inner radius 2$''$ and an outer radius of 3$''$ as a background, we find that a {\sc mekal} and {\sc power-law} model both fit the data equally well. The lack of any obvious non-thermal contribution to the nucleus, even with the new deep 90ks observations confirms that the central nucleus in RX~J1532.9+3021 is radiatively-inefficient \citep[see][]{Hla2011}.

We therefore derive an upper limit to the nuclear $2-10\keV$ flux following the method used in \citet{Hla2011}. The idea is to convert a nuclear count rate into a flux using the {\sc pimms} web interface\footnote[1]{http://heasarc.gsfc.nasa.gov/Tools/w3pimms.html}. More precisely, we compute the total number of counts within the central $2''\times2''$ square region using the $0.5-7\keV$ X-ray image of the new ACIS-S deep observations (ObsID 14009). We then subtract the counts of a surrounding background taken as a square annulus with an inner $3''\times3''$ and an outer $4\times4''$ square, scaled to the same number of pixels. The resulting $3\sigma$ upper limit is then converted to a luminosity with {\sc pimms}, assuming a power-law model with a photon index of 1.9. We find a $3\sigma$ upper limit of $5\times10^{42}\ergps$ in the $2-10$ keV energy range. Note that, changing the location of the nucleus by $\pm2$ pixels does not change the results and that the value is lower than the one determined in \citet{Hla2011}, which is expected since the new observations allow a more precise estimate of the upper limit.

\subsection{Jet powers and X-ray cavities}

There are two clear X-ray cavities associated with the central AGN. The energy stored within each of the cavities can then be estimated from the following equation:

\begin{equation}
E_{\rm bubble}=\frac{\gamma}{\gamma-1}PV \, ,
\label{eq2}
\end{equation}

assuming that the cavities are in pressure equilibrium with the surrounding medium \citep[{\rm e. g.}][]{Bir2004607}. Here, $P$ is the thermal pressure at the radius of the bubble, $V$ is the volume of the cavity and for a relativistic fluid $\gamma$ = 4/3, therefore $E_{\rm bubble}=4PV$. 

We also assume that the cavities are of ellipsoidal shape with $V=4{\pi}R^2_{\rm w}R_{\rm l}/3$, where $R_{\rm l}$ is the projected semi-major axis along the direction of the radio jet axis, and $R_{\rm w}$ is the projected semi-major axis perpendicular to the direction of the radio jet axis. These radii were measured based on the raw, exposure-corrected $0.5-7$ keV image and we assign a 20 per cent uncertainty to $R_{\rm l}$ and $R_{\rm w}$. 

To compute cavity powers, we divide the energy by the bubble age given by the buoyant rise time \citep[][]{Chu2001554} defined as

\begin{equation}
t_\mathrm{buoyant}= R\sqrt{\frac{SC_{\rm D}}{2gV}} \, ,
\label{eq3}
\end{equation}

the refill time \citep[][]{McN2000534} defined as

\begin{equation}
t_\mathrm{refill}= 2\sqrt{\frac{r}{g}} \, ,
\label{eq4}
\end{equation}

or the sound crossing time defined as 

\begin{equation}
t_{c\mathrm{s}}=\frac{R}{c_\mathrm{s}} \, .
\label{eq5}
\end{equation}

Here, $R$ is the projected distance from the optical nucleus to the middle of the cavity. $S$ is the cross-sectional area of the bubble ($S={\pi}R_{\rm w}^2$), $C_{\rm D}$=0.75 is the drag coefficient \citep{Chu2001554}, $g$ is the local gravitational acceleration such that $g=GM(<R)/R^2$, and $r$ is the bubble radius and is equal to $(R_{\rm l}R_{\rm w})^{1/2}$ for an ellipsoidal bubble. Finally, $c_\mathrm{s}$ is the sound crossing time of the gas. Between the three time estimates, it is not clear which is the best to use, although they are not expected to differ significantly. All thermodynamic quantities used for these calculations were taken from the deprojected quantities shown in Fig. \ref{fig11}. The cavity locations are shown in Fig. \ref{fig15}.

Deeper exposures continue to discover new cavities, and in particular older X-ray cavities which have risen buoyantly \citep[e.g.][]{Fab2011418}. Weak shocks and sound waves have been seen to contribute significantly to the power output of the AGN, but these have only been seen through deep observations of nearby systems \citep[e.g.][]{For2005635}. The cavity powers shown in Table \ref{tab4} should therefore be treated as lower limits to the total mechanical energy being injected by the central AGN. We further discuss the cavity powers in Section 7.3.

\section{Discussion}

We have conducted an X-ray (\textit{Chandra} and \textit{XMM}), radio (\textit{VLA} and \textit{GMRT}\footnote[1]{\textit{GMRT} radio contours from \citet{Kal2013}.}) and optical (HST) analysis of the massive, strong cool core cluster RX~J1532.9+3021. We found evidence for powerful mechanical AGN feedback taking place ($P_{\rm cavities}=[22.2\pm8.6]{\times}10^{44}\ergps$), as well as evidence for a cold front that coincides with the edge of the radio mini-halo, and is strongest to the west coincident with the edge of the western X-ray cavity. In the following sections, we discuss various implications of these results. 

\subsection{Cold fronts and metal-enriched outflows}

We first discuss the potential cold front. Cold fronts are contact discontinuities, thought to originate in part from subcluster merger events and subsequent sloshing of low-entropy gas in cluster cores \citep[see a review by][]{Mar2007}. In Section 5, we found evidence for a temperature jump from 4.5 keV to 6.5 keV with increasing radius, strongest to the west at a radius of $\approx$65 kpc (Fig. \ref{fig10}). Fig. \ref{fig11} also showed that there is a pressure break associated with the front, such that the pressure inside the front is low. The right panel of Fig. \ref{fig2} also showed a surface brightness edge at $r\approx$65 kpc, shaped such that it resembles those seen in other cold fronts \citep[see the many examples in][ for comparison]{Mar2007}. All of these results are consistent with the feature being a cold front. 

The front also coincides with the edge of the radio mini-halo. \citet{ZuH2013762} showed that relativistic electrons can be reaccelerated by the turbulence generated from the sloshing of the core. These reaccelerated electrons then produce the mini-halo emission that is coincident with the regions bounded by the sloshing cold fronts \citep[see also][]{Maz2008675}. This could explain the common location of the cold front and edge of the mini-halo in RX~J1532.9+3021, but we note that the cold front is strongest to the west, and at this radius, it also coincides with the edge of the western cavity, similar to what is observed in 4C+55.16 \citep{Hla2011415}. If the cold front is only located to the west, it could be simply due to cool gas being dragged out by the central AGN through the outburst \citep[e.g.][]{Chu2001554}. Alternatively, based on Fig. \ref{fig11}, if one compares the value of the deprojected pressure at a radius of 30 kpc with the one at a radius of 80 kpc, then it seems that the pressure in this region is higher, consistent with what one would expect from a shock front. The break in the pressure profile seen in Fig. \ref{fig11} could therefore also be interpreted as a potential shock front driven by the expansion of the western cavity similar to the high pressure rims seen around the cavities in the Perseus cluster \citep{Fab2003344}.  

The large-scale temperature map in Section 5.1 revealed that the east direction is cooler ($6.0\pm0.2$ keV) than the west ($7.4\pm0.2$ keV). Such an asymmetry in the temperature is most likely caused by a past merger of a cooler sub-cluster with the main cluster. If the core is sloshing and generating the turbulence needed to reaccelerate the electrons in the mini-halo, then a past merger could have caused this. 
\begin{figure}
\centering
\begin{minipage}[c]{0.99\linewidth}
\centering \includegraphics[width=\linewidth]{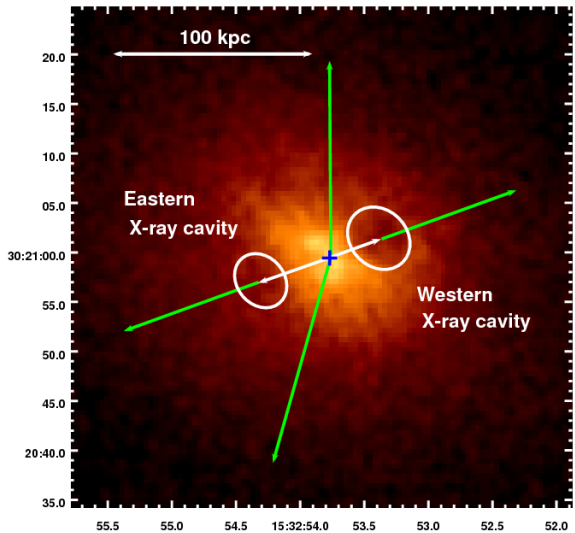}
\end{minipage}
\caption[]{Exposure-corrected $0.5-7\keV$ \textit{Chandra} image, same as in Fig. \ref{fig1}. The X-ray cavities are highlighted with the dimensions used to calculate the powers. We also show the location of the central AGN with the blue cross, taken as the central point source of the \textit{HST} images. The arrows show the direction of the jets, chosen such that they pass through the central AGN and the middle of the X-ray cavities (including the potential ghost cavities to the north and south). The white arrows show the projected distance from the central AGN to the middle of the clear cavity. }
\label{fig15}
\end{figure}

The literature shows several examples of edges which have been interpreted as cold fronts and that have a clear metallicity jump associated with them, but no clear discontinuity in the temperature profile. These include the cold fronts in A2052 \citep[][]{deP2010523}, A2199 \citep{San2006371} and 2A 0335+096 \citep[e.g.][]{Maz2003596}. Others do not show any evidence of a metallicity jump, but clearly show a discontinuity in the temperature profile, such as those in A496 \citep{Dup2003583} and A2204 \citep{San2005356}. In the case of 4C+55.16 \citep{Hla2011415} and M87 \citep{Sim2008482}, there is both a temperature and metallicity jump.

Fig. \ref{fig10} also suggests that there is a metallicity jump to the west at 65 kpc, such that the gas before the front is more metal rich ($\approx0.75~{\rm Z_\odot}$) than gas beyond of the front ($\approx0.40~{\rm Z_\odot}$). The errors bars remain however quite large, and it is not clear from the current data whether the jump is real. Interestingly, there is a second metallicity jump in the western direction at a radius of 40 kpc. This radius coincides with the centre of the western cavity, consistent with the idea that metal rich gas is being dragged out by the cavity, similar to what is observed for other clusters \citep[e.g.]{Kir2011731}. Here, the metallicity before the cavity is $Z\approx0.25~{\rm Z_\odot}$, whereas at 40 kpc, the metalicity is $Z\approx0.75~{\rm Z_\odot}$. Although we do not show the small-scale abundance map in Fig. \ref{fig9} due to the large errors bars, there is an increase in metallicity associated with the western cavity of about $\approx0.5~{\rm Z_\odot}$. If we assume that the cavity has a volume of $V=4{\pi}R^2_{\rm w}R_{\rm l}/3$, and that the electron density at the location of the cavity is approximately 0.05 cm$^{-3}$ (from the deprojected quantities), then the excess of iron mass compared to he iron content before the cavity is around $5\times$10$^7{\rm M_\odot}$. Note that, even more significant enhancements of metal-rich gas ($10^8-10^9M_{\odot}$) along buoyantly rising X-ray cavities have been seen in nearby clusters, such as S{\'e}rsic 159-03 \citep{Wer2011415} and Hydra A \citep{Sim2009493}.

\subsection{The X-ray cavity network}
In \citet{Hla2012421}, we found evidence for only one clear X-ray cavity associated with the central galaxy using the old ACIS-S observations (ObsID 1649). We also found that this cluster had the third strongest cooling luminosity in the 20 MACS clusters with cavities. The only other two stronger cool core clusters were MACS~J0947.2+7623 and MACS~J1447.4+0827, with cooling luminosities 20 to 40 per cent larger respectively. Based only on the western cavity, we concluded that the power stored within the cavity fell short by a factor of $\approx2-3$ of the cooling luminosity. 

With the new 90ks observations, we find evidence for at least two X-ray cavities. We confirm the presence of the western cavity, and report on the newly discovered eastern cavity located at a similar radius and of similar size (see Fig. \ref{fig15}). Both cavities also coincident with radio jets as seen from the 1.4 GHz VLA image (see left panel of Fig. \ref{fig1b}). The direction of the radio jets and the central AGN are also well aligned with each other.

\begin{figure}
\centering
\begin{minipage}[c]{0.99\linewidth}
\centering \includegraphics[width=\linewidth]{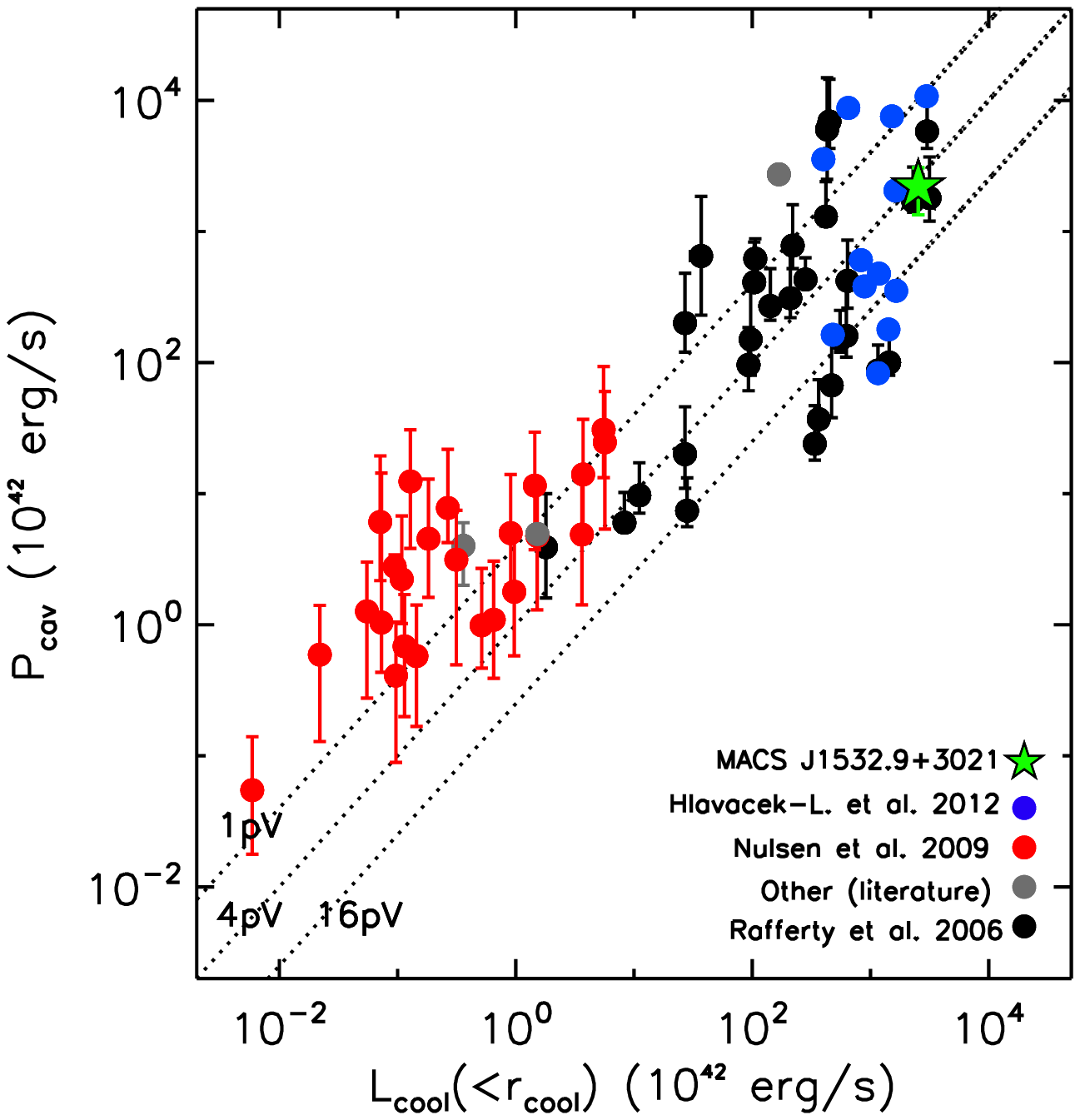}
\end{minipage}
\caption[]{Cavity power as a function of cooling luminosity. Shown are RX~J1532.9+3021 (star symbol), as well as the data points of \citet{Hla2012421}, \citet{Raf2006652}, \citet{Nul2009} and in grey, the data points of \citet{Dun2008385}, \citet{Ran2009700} and \citet{Jet2008384}. Figure adapted from \citet{Fab2012}. }
\label{fig14}
\end{figure}

The total enthalpy associated with both X-ray cavities ($4PV$) is around $3{\times}10^{60}\erg$ and the total power is $(22.2\pm8.6){\times}10^{44}\ergps$. This is sufficient to prevent the gas from cooling where $L_{\rm cool}\approx25\times10^{44}\ergps$ in the $0.01-50.0$ keV range. We show this more clearly in Fig. \ref{fig14}. Beyond this energy range, the plasma model does not contribute significantly, and this value of $L_{\rm cool}$ represents the total cooling luminosity.

In comparison, the X-ray cavities in RX~J1532.9+3021 are around an order of magnitude more powerful than the inner X-ray cavities of the Perseus cluster \citep[$z=0.018$;][]{Dun2004355} and M87 \citep[$z=0.0044$;][]{For2005635}, and several times that of Hydra A \citep[$z=0.0538$;][]{Nul2005628}. They are also around two times more powerful than those in Cygnus A ($z=0.056$), Abell 1835 ($z=0.253$) and PKS 0745-19 \citep[$z=0.01028$;][]{Raf2006652}, as well as MACS~J1423.8+2404 \citep[$z=0.5449$;][]{Hla2012421}. These X-ray cavities belong to the extreme realm of AGN feedback, similar to those in MACS~J0947.2+7623 \citep[$z=0.354$;][]{Cav2011732} and MACS~J1931.8-2634 \citep[$z=0.352$;][]{Ehl2011411}, although not as powerful as those in MS 0735.6+7421 \citep[$z=0.216$;][]{McN2005433}.

The right panel of Fig. \ref{fig4} reveals hints of a large northern depression, adjacent to the faint northern shell of the western cavity. This could be due to the western cavity leaking out into the intracluster medium. The leaked relativistic particles could then contribute to the heating \citep[e.g.][]{Guo2008384}, as well as account for the radio mini-halo. Such a phenomenon has been seen in M84 \citep{Fin2008686}, as well as in Abell 2052 for the outermost shells of the southern and northern X-ray cavities. The latter appear to be breaking apart and radio plasma is seen to leak out, but only slightly beyond the cavities \citep{Bla2011737}. It is therefore difficult to explain why the leakage in RX~J1532.9+3021 extends to such a large scale. Due to the presence of the radio mini-halo, it is also not clear from the radio observations if leakage is occurring. 

As mentioned in Section 4, this potential large northern depression could also represent a single older X-ray cavity, or it could represent the accumulation of past western outflows along the northern direction. The surface brightness decrement associated with the depression is on the order of 10 to 15 per cent (see Fig. \ref{fig5}), whereas the two clear eastern and western cavities have a decrement of 30 per cent. The exact shape of the depression is therefore less well defined. Taking this into account, and considering that projection effects may be important, we assign a radius of $15\pm7$ kpc to the northern depression, where the large uncertainty attempts to account for the unknown factors. Considering that it has roughly the same size as the western cavity, then it is unlikely that the depression represents the accumulation of past western outflows since one would expect it to be significantly larger. If the northern cavity is largely elongated along our line of sight, then the volume could be significantly larger, but this remains unclear. Considering the more likely scenario that it represents one single older outburst, then the power associated with it is $\approx5-20{\times}10^{44}\ergps$. However, it is important to note that the projected distance from the central AGN to this northern depression is the same as for the western X-ray cavity. Although strong projection effects can account for this, if it were an older X-ray cavity, one would expect it to be located beyond the western X-ray cavity.

Interestingly, the southern optical emission filaments extend beyond the eastern cavity, out to 50 kpc in radius (Fig. \ref{fig6b}). Some studies suggest that filaments form from cooling of the hot ICM \citep[e.g.][]{McD2010721}, while others argue that some filaments originate from cool gas being dragged out in the wake of a cavity \citep[e.g.][]{Chu2013}. The southern filaments may therefore indicate that the past outbursts were aligned along the north to south direction. Even more, the 325 MHz radio map shown in the right panel of Fig. \ref{fig1b} shows a diffuse north to south jet-like feature not clearly seen at 610 MHz or 1.4 GHz, consistent with particle aging. 

Hence, based on the existence of these southern optical filaments, as well as the north to south 325 MHz jet-like structure, we conclude that RX~J1532.9+3021 most likely harbours older AGN-driven outflows along the north to south direction. This implies that either sloshing effects are important and have caused the outflows to change directions, or that the jet is precessing (e.g. Dunn et al. 2006, Sternberg \& Soker 2008, Canning et al. 2013)\nocite{Dun2006366,Ste2008384,Can2013}. Furthermore, the potential northern X-ray depression lies along this direction, and may therefore represent the wake of the northern older outflow no longer seen at high-frequency radio wavelengths, known as a ghost cavity. Ghost cavities have been seen in other systems, although at much lower redshifts. 

If the jet is precessing, then the difference in angle between the location of the current outburst and that of the older outburst is $\approx$70$^o$ for the north/west and $\approx50^o$ for the south/east (Fig. \ref{fig15}). Assuming that the cavities are rising in the plane of the sky, which is probably not the case, we estimate from the difference in buoyancy rise time that the jet is precessing at $\approx{20}^o$ per $10^7$ yrs. This is significantly larger than what Canning et al. (2013) find for Abell 3581, but less than what \citet{Dun2006366} find for the Perseus cluster. For the latter, the authors find a best-fitting solution consisting of a precession axis inclined by 120$^o$ with respect to the line of sight, with an opening angle of 50$^o$ and a precession timescale of $\approx3\times10^7$ yrs.

\begin{figure}
\centering
\begin{minipage}[c]{0.99\linewidth}
\centering \includegraphics[width=\linewidth]{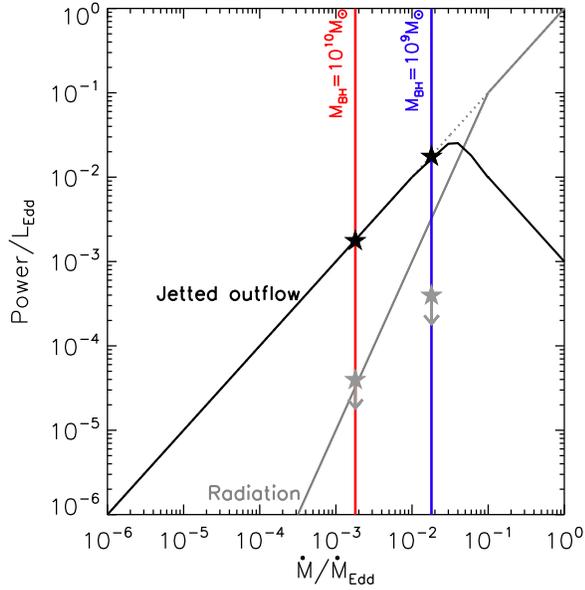}
\end{minipage}
\caption[]{Sketch of the power emerging from a black hole as a function of the accretion rate, both in terms of the Eddington ratio \citep[adapted from][]{Chu2005363}. We illustrate the approximate position of RX~J1532.9+3021 in this plot, in terms of the mechanical (black stars) and radiative (grey stars) powers assuming a $10^{9}M_{\odot}$ black hole (blue line) and a $10^{10}M_{\odot}$ black hole (red line). The observed radiative power (grey star) for a $10^{9}M_{\odot}$ black hole appears to lie below the predicted line, unlike the value for a $10^{10}M_{\odot}$ black hole.}
\label{fig16}
\end{figure}

\subsection{Powering the central AGN}

Assuming that the X-ray cavities were powered by the release of gravitational binding energy from accretion onto the central black hole, we can estimate the minimum amount of mass the black hole must have accreted to power the outburst with the following equation:

\begin{equation}
\Delta M_\mathrm{BH}= \frac{E_\mathrm{cavities}}{\eta c^2}          \, .
\label{eq6}
\end{equation}

Here, the accretion efficiency ($\eta$) is assumed to be equal to 0.1. We find a total mass gain of $2\times10^7M_\mathrm{\odot}$ if we only consider the two clear X-ray cavities, but this amount goes up to $\approx{0.5}\times10^8M_\mathrm{\odot}$ if we also consider the potential ghost cavities. For the two inner X-ray cavities, this translate to an average accretion rate of $0.4M_\mathrm{\odot}$ yr$^{-1}$ over the buoyancy rise time of the cavities.

There are no mass estimates of the black hole available in the literature. The BCG is included only in the 2MASS point source catalogue, but the quoted magnitude significantly underestimates the total flux since the point spread function extends only to $\approx2''$ ($\approx$10 kpc) and the total envelope of the BCG extends to some 50 kpc in radius based on the near-infrared HST observations. There are also no {\sc hyperleda} stellar velocity dispersion measurements. For the purposes of this study, we therefore assume that the black hole mass lies within $10^{9-10}M_{\odot}$ based on the known black hole masses of other BCGs from modelling \citep[e.g.][]{McC2011480}. Assuming that the Eddington luminosity is defined as
\begin{equation}
\frac{L_\mathrm{Edd}}{\rm{\ergps}} = 1.26 \times 10^{47} \left(\frac{M_\mathrm{BH}}{10^9M_{\odot}}\right)  \, 
\label{eq7}
\end{equation}

for a fully ionized plasma, the mechanical power emerging from RX~J1532.9+3021 in the form of the two clear X-ray cavities corresponds to $2$ per cent of the Eddington luminosity for a $10^{9}M_{\odot}$ black hole and $0.2$ per cent of the Eddington luminosity for a $10^{10}M_{\odot}$ black hole. 

Such a high Eddington fraction provides some constraints on the possible fuelling mechanisms of the black hole. Since BCGs are embedded in hot atmospheres, one appealing solution is fuelling from the surrounding hot halo \citep[e.g.][]{All2006372}. This type of accretion, known as Bondi accretion \citep{Bon1952112}, involves a spherical inflow of the hot X-ray emitting ga. For a given atmosphere temperature of $T$ and density of $n_e$, assuming an adiabatic index of $\Gamma=5/3$, it is given by the following equation: 

\begin{equation}
\frac{\dot{M}_\mathrm{B}}{M_{\odot} {\mathrm{yr}}^{-1}} = 0.012 \left(\frac{k_{B}T}{\mathrm{keV}}\right)^{-3/2}   \left(\frac{n_e}{\mathrm{cm^{-3}}}\right)   \left(\frac{M_\mathrm{BH}}{10^9M_{\odot}}\right)^2 \, .
\label{eq8}
\end{equation}

To compute the true Bondi accretion rate, we would need to resolve the Bondi radius which is on the order of hundreds of pc in RX~J1532.9+3021. This is not possible with current X-ray observatories. We therefore only compute a rough estimate of the Bondi accretion rate by considering the temperature and density at the Bondi radius found by \citet{Rus2013} for the most powerful outflow in their subsample of 13 systems with X-ray cavities where they were able to obtain a resolution close to the Bondi radius, NGC 507 ($P_{\rm cavities}\approx2\times10^{43}\ergps$). The latter has an approximate temperature at the Bondi radius of $\approx0.5$ keV and an electron of density $0.8$ cm$^{-3}$. This corresponds to a Bondi accretion rate of $0.03M_{\odot}$ yr$^{-1}$ for a $10^{9}M_{\odot}$ black hole or $3.0M_{\odot}$ yr$^{-1}$ for a $10^{10}M_{\odot}$. If we consider the temperature and density values at the Bondi radius of \citet{All2006372} for NGC 507, then the accretion rate increases by a factor of two. It is therefore currently difficult to explain the powerful cavities in RX~J1532.9+3021 through Bondi accretion for a $10^{9}M_{\odot}$ black hole, although the density within the Bondi radius may be even larger than that in NGC 507 considering the extreme cool core properties of the cluster. Interestingly, the BCG in RX~J1532.9+3021 has one of the most massive molecular gas detections known \citet[][$2.5\times10^{11}M_{\odot}$]{Edg2001328}. The extensive molecular reservoir surrounding the BCG is therefore significantly larger than the mass required to power the outbursts, making cold accretion also a viable solution to the fuelling mechanism of the AGN.

In Section 6.1, we derived a $3\sigma$ upper limit of $5\times10^{42}\ergps$ for the $2-10$ keV nuclear luminosity of the central AGN. Radiatively-inefficient AGN, with luminosities of $<1$ per cent Eddington, often lack the optical/UV big blue bump normally seen in more radiatively-efficient AGN. The value for the bolometric conversion factor is also not well known for such low-luminosity AGN. \citet{Vas2007381} find a typical correction factor of 10 for radiatively-inefficient AGN, whereas \citet{Mer2007381} and \citet{Rus2013} both adopt a correction factor of 5 for the AGN in BCGs. Here we adopt a factor of 10, but stress that if we were to use a factor of 5 then our following conclusions would be stronger. Applying this bolometric correction, we find that the radiative output from the central AGN is less than 0.04 per cent Eddginton ($M_{\rm BH}=10^{9}M_{\odot}$) or 0.004 per cent Eddington ($M_{\rm BH}=10^{10}M_{\odot}$). 

Many authors have argued that black holes typically transit between two major states \citep[e.g.][]{Gal2003344}. The first, known as the low-hard state, is characterized by a radiatively-inefficient black hole that is accreting at low rates ($<1$ per cent ${\rm \dot{M}_{\rm Edd}}$), yet capable of driving powerful jetted outflows. The second, known as the high-soft state, is a thermal state characterized by an efficiently accreting black hole ($>1$ per cent ${\rm \dot{M}_{\rm Edd}}$) in which the jet is quenched.

These two states are illustrated in Fig. \ref{fig16}, adopted from \citet{Chu2005363}. We stress that Fig. \ref{fig16} is a sketch, and therefore illustrates only qualitatively the properties of the two states. We highlight the approximate position of of RX~J1532.9+3021 in the figure, by computing the total accretion rate ($\dot{M}$) as the sum required to power the X-ray cavities (for the two clear X-ray cavities) and the bolometric radiative luminosity. We then divide this quantity by the Eddington luminosity for a $10^{9}M_{\odot}$ black hole (blue) and a $10^{10}M_{\odot}$ black hole (red). Although this figure is approximate, it shows nicely that a $10^{9}M_{\odot}$ black hole would need to be accreting at very high rates to explain the powerful jets, so much that, the power emerging from the central AGN in terms of radiation is expected to be on the same level as the mechanical power. This is clearly not seen. If the central black hole has a mass of $10^{10}M_{\odot}$, then the radiative-inefficiency of the central AGN can be easily explained by considering that the AGN is in the low-hard state.

Finally, we mention the possibility that the central AGN may be highly-obscured. The power then needs to re-emerge in the infrared. Interestingly, RX~J1532.9+3021 has the fourth most infrared luminous BCG in the 62 BCG sample of \citet[][$\L_{\rm IR}=22.6\times10^{44}\ergps$]{Qui2008176}. Furthermore, \citet{Gan2009502} argued that the intrinsic $2-10\keV$ X-ray luminosity of an AGN (corrected for absorption) is correlated with its mid-infrared luminosity at $12\mu{\rm m}$. If this correlation still holds for BCGs, then by determining a rough estimate of the $12\mu{\rm m}$ luminosity from the $8\mu{\rm m}$ and $24\mu{\rm m}$ measurements of \citet[][; $\approx10^{44}\ergps$]{Qui2008176}, the corresponding intrinsic $2-10\keV$ X-ray luminosity should be $\approx10^{44}\ergps$. Since we only measure an upper limit of $5\times10^{42}\ergps$, this would imply that the source is highly-obscured. To obtain such a strong attenuation (factor of $>20$), the absorbing column density would need to be around $10^{24}$ cm$^{-2}$. If the true luminosity is even higher ($\approx{5}\times10^{44}\ergps$), then the absorbing column density would need to be several $10^{24}$ cm$^{-2}$, making the source Compton-thick. However, we note that the [O\thinspace{\sc iii}]$\lambda5007$/$H\beta$ line ratio, based on the line measurements of \citet{Cra1999306} using an aperture of $\approx$10 kpc, is very low ($0.5-0.6$). The infrared emission arising from the BCG is therefore more likely dominated by star formation than an AGN \citep[e.g.][]{Qui2008176,ODe2008681}. 

\citet{ODe2008681} estimate that the SFR, based on the Spitzer infrared data is $\approx$100 $M_{\odot}$ yr$^{-1}$. We find a consistent UV SFR of $76\pm$38 $M_{\odot}$ yr$^{-1}$. The heating from the outflows most likely proceeds in a volume averaged way and is sufficient to offset the cooling on average. However, locally, some cooling in the form of the filaments may occur due local instabilities. In the case of MACS~J1931.8-2634, star formation is also occurring even though the cavity power exceeds the X-ray cooling luminosity. This disagrees with the empirical star formation conditions found in \citet{Raf2008687} suggesting that star formation only occurs in the cavity power is less than the cooling luminosity.

Note that many of the calculations in this section rely on strong assumptions, such as the accretion efficiency $\eta$. The latter is a strong function of the inner boundary conditions of the accretion flow and can vary from 0.057 for a non-spinning Schwarzschild black hole to 0.42 for a maximally-spinning Kerr black hole \citep{Nov1973}. A rapidly spinning black hole is therefore also an attractive solution to explain the high jet powers observed in RX~J1532.9+3021 and the radiative-inefficiency of the central AGN \citep[e.g][]{Tch2011418}.

\acknowledgments
JHL is supported by NASA through the Einstein Fellowship Program, grant number PF2-130094. ACF thanks the Royal Society. AVDL thanks the HST award HST-AR-12654.01-A. G.B Taylor and C.K. Grimes acknowledge support from NASA through Chandra Award number GO2-13149X issued by the Chandra X-ray Observatory Center, which is operated by the Smithsonian Astrophysical Observatory for and on behalf of NASA under contract NAS8-03060. JHL also thanks D. J. Walton for helpful skype discussions about \textit{XMM} data, and the referee for his or her suggestions.

\bibliographystyle{mn2e}
\bibliography{bibli}

\begin{thebibliography}{}

\bibitem[\protect\citeauthoryear{{Allen}, {Dunn}, {Fabian}, {Taylor} \&
  {Reynolds}}{{Allen} et~al.}{2006}]{All2006372}
{Allen} S.~W.,  {Dunn} R.~J.~H.,  {Fabian} A.~C.,  {Taylor} G.~B.,
  {Reynolds} C.~S.,  2006, \mnras, 372, 21

\bibitem[\protect\citeauthoryear{{Allen} \& {Fabian}}{{Allen} \&
  {Fabian}}{1997}]{All1997286}
{Allen} S.~W.,  {Fabian} A.~C.,  1997, \mnras, 286, 583

\bibitem[\protect\citeauthoryear{{Anders} \& {Grevesse}}{{Anders} \&
  {Grevesse}}{1989}]{And198953}
{Anders} E.,  {Grevesse} N.,  1989, \gca, 53, 197

\bibitem[\protect\citeauthoryear{{B{\^i}rzan}, {McNamara}, {Nulsen}, {Carilli}
  \& {Wise}}{{B{\^i}rzan} et~al.}{2008}]{Bir2008686}
{B{\^i}rzan} L.,  {McNamara} B.~R.,  {Nulsen} P.~E.~J.,  {Carilli} C.~L.,
  {Wise} M.~W.,  2008, \apj, 686, 859

\bibitem[\protect\citeauthoryear{{B{\^i}rzan}, {Rafferty}, {McNamara}, {Wise}
  \& {Nulsen}}{{B{\^i}rzan} et~al.}{2004}]{Bir2004607}
{B{\^i}rzan} L.,  {Rafferty} D.~A.,  {McNamara} B.~R.,  {Wise} M.~W.,
  {Nulsen} P.~E.~J.,  2004, \apj, 607, 800

\bibitem[\protect\citeauthoryear{{Blanton}, {Randall}, {Clarke}, {Sarazin},
  {McNamara}, {Douglass} \& {McDonald}}{{Blanton} et~al.}{2011}]{Bla2011737}
{Blanton} E.~L.,  {Randall} S.~W.,  {Clarke} T.~E.,  {Sarazin} C.~L.,
  {McNamara} B.~R.,  {Douglass} E.~M.,    {McDonald} M.,  2011, \apj, 737, 99

\bibitem[\protect\citeauthoryear{{Bondi}}{{Bondi}}{1952}]{Bon1952112}
{Bondi} H.,  1952, \mnras, 112, 195

\bibitem[\protect\citeauthoryear{{Cardelli}, {Clayton} \& {Mathis}}{{Cardelli}
  et~al.}{1989}]{Car1989345}
{Cardelli} J.~A.,  {Clayton} G.~C.,    {Mathis} J.~S.,  1989, \apj, 345, 245

\bibitem[\protect\citeauthoryear{{Cavagnolo}, {McNamara}, {Nulsen}, {Carilli},
  {Jones} \& {B{\^i}rzan}}{{Cavagnolo} et~al.}{2010}]{Cav2010720}
{Cavagnolo} K.~W.,  {McNamara} B.~R.,  {Nulsen} P.~E.~J.,  {Carilli} C.~L.,
  {Jones} C.,    {B{\^i}rzan} L.,  2010, \apj, 720, 1066

\bibitem[\protect\citeauthoryear{{Cavagnolo}, {McNamara}, {Wise}, {Nulsen},
  {Br{\"u}ggen}, {Gitti} \& {Rafferty}}{{Cavagnolo} et~al.}{2011}]{Cav2011732}
{Cavagnolo} K.~W.,  {McNamara} B.~R.,  {Wise} M.~W.,  {Nulsen} P.~E.~J.,
  {Br{\"u}ggen} M.,  {Gitti} M.,    {Rafferty} D.~A.,  2011, \apj, 732, 71

\bibitem[\protect\citeauthoryear{{Churazov}, {Br{\"u}ggen}, {Kaiser},
  {B{\"o}hringer} \& {Forman}}{{Churazov} et~al.}{2001}]{Chu2001554}
{Churazov} E.,  {Br{\"u}ggen} M.,  {Kaiser} C.~R.,  {B{\"o}hringer} H.,
  {Forman} W.,  2001, \apj, 554, 261

\bibitem[\protect\citeauthoryear{{Churazov}, {Ruszkowski} \&
  {Schekochihin}}{{Churazov} et~al.}{2013}]{Chu2013}
{Churazov} E.,  {Ruszkowski} M.,    {Schekochihin} A.,  2013, ArXiv e-prints

\bibitem[\protect\citeauthoryear{{Churazov}, {Sazonov}, {Sunyaev}, {Forman},
  {Jones} \& {B{\"o}hringer}}{{Churazov} et~al.}{2005}]{Chu2005363}
{Churazov} E.,  {Sazonov} S.,  {Sunyaev} R.,  {Forman} W.,  {Jones} C.,
  {B{\"o}hringer} H.,  2005, \mnras, 363, L91

\bibitem[\protect\citeauthoryear{{Crawford}, {Allen}, {Ebeling}, {Edge} \&
  {Fabian}}{{Crawford} et~al.}{1999}]{Cra1999306}
{Crawford} C.~S.,  {Allen} S.~W.,  {Ebeling} H.,  {Edge} A.~C.,    {Fabian}
  A.~C.,  1999, \mnras, 306, 857

\bibitem[\protect\citeauthoryear{{de Plaa}, {Werner}, {Simionescu}, {Kaastra},
  {Grange} \& {Vink}}{{de Plaa} et~al.}{2010}]{deP2010523}
{de Plaa} J.,  {Werner} N.,  {Simionescu} A.,  {Kaastra} J.~S.,  {Grange}
  Y.~G.,    {Vink} J.,  2010, \aap, 523, A81+

\bibitem[\protect\citeauthoryear{{Dong}, {Rasmussen} \& {Mulchaey}}{{Dong}
  et~al.}{2010}]{Don2010712}
{Dong} R.,  {Rasmussen} J.,    {Mulchaey} J.~S.,  2010, \apj, 712, 883

\bibitem[\protect\citeauthoryear{{Dunn} \& {Fabian}}{{Dunn} \&
  {Fabian}}{2004}]{Dun2004355}
{Dunn} R.~J.~H.,  {Fabian} A.~C.,  2004, \mnras, 355, 862

\bibitem[\protect\citeauthoryear{{Dunn} \& {Fabian}}{{Dunn} \&
  {Fabian}}{2006}]{Dun2006373}
{Dunn} R.~J.~H.,  {Fabian} A.~C.,  2006, \mnras, 373, 959

\bibitem[\protect\citeauthoryear{{Dunn} \& {Fabian}}{{Dunn} \&
  {Fabian}}{2008}]{Dun2008385}
{Dunn} R.~J.~H.,  {Fabian} A.~C.,  2008, \mnras, 385, 757

\bibitem[\protect\citeauthoryear{{Dunn}, {Fabian} \& {Sanders}}{{Dunn}
  et~al.}{2006}]{Dun2006366}
{Dunn} R.~J.~H.,  {Fabian} A.~C.,    {Sanders} J.~S.,  2006, \mnras, 366, 758

\bibitem[\protect\citeauthoryear{{Dupke} \& {White} III}{{Dupke} \&
  {White}}{2003}]{Dup2003583}
{Dupke} R.,  {White} III R.~E.,  2003, \apjl, 583, L13

\bibitem[\protect\citeauthoryear{{Ebeling}, {Edge}, {Mantz}, {Barrett},
  {Henry}, {Ma} \& {van Speybroeck}}{{Ebeling} et~al.}{2010}]{Ebe2010407}
{Ebeling} H.,  {Edge} A.~C.,  {Mantz} A.,  {Barrett} E.,  {Henry} J.~P.,  {Ma}
  C.~J.,    {van Speybroeck} L.,  2010, \mnras, 407, 83

\bibitem[\protect\citeauthoryear{{Edge}}{{Edge}}{2001}]{Edg2001328}
{Edge} A.~C.,  2001, \mnras, 328, 762

\bibitem[\protect\citeauthoryear{{Edge}, {Stewart} \& {Fabian}}{{Edge}
  et~al.}{1992}]{Edg1992258}
{Edge} A.~C.,  {Stewart} G.~C.,    {Fabian} A.~C.,  1992, \mnras, 258, 177

\bibitem[\protect\citeauthoryear{{Ehlert}, {Allen}, {von der Linden},
  {Simionescu}, {Werner}, {Taylor}, {Gentile}, {Ebeling}, {Allen}, {Applegate},
  {Dunn}, {Fabian}, {Kelly}, {Million}, {Morris}, {Sanders} \&
  {Schmidt}}{{Ehlert} et~al.}{2011}]{Ehl2011411}
{Ehlert} S.,  {Allen} S.~W.,  {von der Linden} A.,  {Simionescu} A.,  {Werner}
  N.,  {Taylor} G.~B.,  {Gentile} G.,  {Ebeling} H.,  {Allen} M.~T.,
  {Applegate} D.,  {Dunn} R.~J.~H.,  {Fabian} A.~C.,  {Kelly} P.,  {Million}
  E.~T.,  {Morris} R.~G.,  {Sanders} J.~S.,    {Schmidt} R.~W.,  2011, \mnras,
  411, 1641

\bibitem[\protect\citeauthoryear{{Fabian}}{{Fabian}}{2012}]{Fab2012}
{Fabian} A.~C.,  2012, \araa, 50, 455

\bibitem[\protect\citeauthoryear{{Fabian}, {Sanders}, {Allen}, {Canning},
  {Churazov}, {Crawford}, {Forman}, {Gabany}, {Hlavacek-Larrondo}, {Johnstone},
  {Russell}, {Reynolds}, {Salom{\'e}}, {Taylor} \& {Young}}{{Fabian}
  et~al.}{2011}]{Fab2011418}
{Fabian} A.~C.,  {Sanders} J.~S.,  {Allen} S.~W.,  {Canning} R.~E.~A.,
  {Churazov} E.,  {Crawford} C.~S.,  {Forman} W.,  {Gabany} J.,
  {Hlavacek-Larrondo} J.,  {Johnstone} R.~M.,  {Russell} H.~R.,  {Reynolds}
  C.~S.,  {Salom{\'e}} P.,  {Taylor} G.~B.,    {Young} A.~J.,  2011, \mnras,
  418, 2154

\bibitem[\protect\citeauthoryear{{Fabian}, {Sanders}, {Allen}, {Crawford},
  {Iwasawa}, {Johnstone}, {Schmidt} \& {Taylor}}{{Fabian}
  et~al.}{2003}]{Fab2003344}
{Fabian} A.~C.,  {Sanders} J.~S.,  {Allen} S.~W.,  {Crawford} C.~S.,  {Iwasawa}
  K.,  {Johnstone} R.~M.,  {Schmidt} R.~W.,    {Taylor} G.~B.,  2003, \mnras,
  344, L43

\bibitem[\protect\citeauthoryear{{Falcke}, {Rieke}, {Rieke}, {Simpson} \&
  {Wilson}}{{Falcke} et~al.}{1998}]{Fal1998494}
{Falcke} H.,  {Rieke} M.~J.,  {Rieke} G.~H.,  {Simpson} C.,    {Wilson} A.~S.,
  1998, \apjl, 494, L155

\bibitem[\protect\citeauthoryear{{Ferrari}, {Govoni}, {Schindler}, {Bykov} \&
  {Rephaeli}}{{Ferrari} et~al.}{2008}]{Fer2008134}
{Ferrari} C.,  {Govoni} F.,  {Schindler} S.,  {Bykov} A.~M.,    {Rephaeli} Y.,
  2008, \ssr, 134, 93

\bibitem[\protect\citeauthoryear{{Finoguenov}, {Ruszkowski}, {Jones},
  {Br{\"u}ggen}, {Vikhlinin} \& {Mandel}}{{Finoguenov}
  et~al.}{2008}]{Fin2008686}
{Finoguenov} A.,  {Ruszkowski} M.,  {Jones} C.,  {Br{\"u}ggen} M.,  {Vikhlinin}
  A.,    {Mandel} E.,  2008, \apj, 686, 911

\bibitem[\protect\citeauthoryear{{Forman}, {Nulsen}, {Heinz}, {Owen}, {Eilek},
  {Vikhlinin}, {Markevitch}, {Kraft}, {Churazov} \& {Jones}}{{Forman}
  et~al.}{2005}]{For2005635}
{Forman} W.,  {Nulsen} P.,  {Heinz} S.,  {Owen} F.,  {Eilek} J.,  {Vikhlinin}
  A.,  {Markevitch} M.,  {Kraft} R.,  {Churazov} E.,    {Jones} C.,  2005,
  \apj, 635, 894

\bibitem[\protect\citeauthoryear{{Gallo}, {Fender} \& {Pooley}}{{Gallo}
  et~al.}{2003}]{Gal2003344}
{Gallo} E.,  {Fender} R.~P.,    {Pooley} G.~G.,  2003, \mnras, 344, 60

\bibitem[\protect\citeauthoryear{{Gandhi}, {Horst}, {Smette}, {H{\"o}nig},
  {Comastri}, {Gilli}, {Vignali} \& {Duschl}}{{Gandhi}
  et~al.}{2009}]{Gan2009502}
{Gandhi} P.,  {Horst} H.,  {Smette} A.,  {H{\"o}nig} S.,  {Comastri} A.,
  {Gilli} R.,  {Vignali} C.,    {Duschl} W.,  2009, \aap, 502, 457

\bibitem[\protect\citeauthoryear{Giacintucci \& {et al.}}{Giacintucci \& {et
  al.}}{2013}]{Gia2013}
Giacintucci S.,  {et al.} 2013, \apj~submitted

\bibitem[\protect\citeauthoryear{{Greisen}}{{Greisen}}{2003}]{Gre2003285}
{Greisen} E.~W.,  2003, Information Handling in Astronomy - Historical Vistas,
  285, 109

\bibitem[\protect\citeauthoryear{{Guo} \& {Oh}}{{Guo} \&
  {Oh}}{2008}]{Guo2008384}
{Guo} F.,  {Oh} S.~P.,  2008, \mnras, 384, 251

\bibitem[\protect\citeauthoryear{{G{\"u}ver} \& {{\"O}zel}}{{G{\"u}ver} \&
  {{\"O}zel}}{2009}]{Guv2010400}
{G{\"u}ver} T.,  {{\"O}zel} F.,  2009, \mnras, 400, 2050

\bibitem[\protect\citeauthoryear{{Hlavacek-Larrondo} \&
  {Fabian}}{{Hlavacek-Larrondo} \& {Fabian}}{2011}]{Hla2011}
{Hlavacek-Larrondo} J.,  {Fabian} A.~C.,  2011, \mnras, 413, 313

\bibitem[\protect\citeauthoryear{{Hlavacek-Larrondo}, {Fabian}, {Edge},
  {Ebeling}, {Allen}, {Sanders} \& {Taylor}}{{Hlavacek-Larrondo}
  et~al.}{2013}]{Hla2013431}
{Hlavacek-Larrondo} J.,  {Fabian} A.~C.,  {Edge} A.~C.,  {Ebeling} H.,  {Allen}
  S.~W.,  {Sanders} J.~S.,    {Taylor} G.~B.,  2013, \mnras, 431, 1638

\bibitem[\protect\citeauthoryear{{Hlavacek-Larrondo}, {Fabian}, {Edge},
  {Ebeling}, {Sanders}, {Hogan} \& {Taylor}}{{Hlavacek-Larrondo}
  et~al.}{2012}]{Hla2012421}
{Hlavacek-Larrondo} J.,  {Fabian} A.~C.,  {Edge} A.~C.,  {Ebeling} H.,
  {Sanders} J.~S.,  {Hogan} M.~T.,    {Taylor} G.~B.,  2012, \mnras, 421, 1360

\bibitem[\protect\citeauthoryear{{Hlavacek-Larrondo}, {Fabian}, {Sanders} \&
  {Taylor}}{{Hlavacek-Larrondo} et~al.}{2011}]{Hla2011415}
{Hlavacek-Larrondo} J.,  {Fabian} A.~C.,  {Sanders} J.~S.,    {Taylor} G.~B.,
  2011, \mnras, 415, 3520

\bibitem[\protect\citeauthoryear{{Jetha}, {Hardcastle}, {Babul}, {O'Sullivan},
  {Ponman}, {Raychaudhury} \& {Vrtilek}}{{Jetha} et~al.}{2008}]{Jet2008384}
{Jetha} N.~N.,  {Hardcastle} M.~J.,  {Babul} A.,  {O'Sullivan} E.,  {Ponman}
  T.~J.,  {Raychaudhury} S.,    {Vrtilek} J.,  2008, \mnras, 384, 1344

\bibitem[\protect\citeauthoryear{{Johnstone}, {Fabian} \& {Nulsen}}{{Johnstone}
  et~al.}{1987}]{Joh1987224}
{Johnstone} R.~M.,  {Fabian} A.~C.,    {Nulsen} P.~E.~J.,  1987, \mnras, 224,
  75

\bibitem[\protect\citeauthoryear{{Kalberla}, {Burton}, {Hartmann}, {Arnal},
  {Bajaja}, {Morras} \& {P{\"o}ppel}}{{Kalberla} et~al.}{2005}]{Kal2005440}
{Kalberla} P.~M.~W.,  {Burton} W.~B.,  {Hartmann} D.,  {Arnal} E.~M.,  {Bajaja}
  E.,  {Morras} R.,    {P{\"o}ppel} W.~G.~L.,  2005, \aap, 440, 775

\bibitem[\protect\citeauthoryear{{Kale}, {Venturi}, {Giacintucci}, {Dallacasa},
  {Cassano}, {Brunetti}, {Macario} \& {Athreya}}{{Kale} et~al.}{2013}]{Kal2013}
{Kale} R.,  {Venturi} T.,  {Giacintucci} S.,  {Dallacasa} D.,  {Cassano} R.,
  {Brunetti} G.,  {Macario} G.,    {Athreya} R.,  2013, ArXiv e-prints

\bibitem[\protect\citeauthoryear{{Kennicutt}
  Jr.}{{Kennicutt}}{1998}]{Ken1998498}
{Kennicutt} Jr. R.~C.,  1998, \apj, 498, 541

\bibitem[\protect\citeauthoryear{{Kirkpatrick}, {McNamara} \&
  {Cavagnolo}}{{Kirkpatrick} et~al.}{2011}]{Kir2011731}
{Kirkpatrick} C.~C.,  {McNamara} B.~R.,    {Cavagnolo} K.~W.,  2011, \apjl,
  731, L23

\bibitem[\protect\citeauthoryear{{Mantz}, {Allen}, {Ebeling}, {Rapetti} \&
  {Drlica-Wagner}}{{Mantz} et~al.}{2010}]{Man2010406}
{Mantz} A.,  {Allen} S.~W.,  {Ebeling} H.,  {Rapetti} D.,    {Drlica-Wagner}
  A.,  2010, \mnras, 406, 1773

\bibitem[\protect\citeauthoryear{{Markevitch} \& {Vikhlinin}}{{Markevitch} \&
  {Vikhlinin}}{2007}]{Mar2007}
{Markevitch} M.,  {Vikhlinin} A.,  2007, \physrep, 443, 1

\bibitem[\protect\citeauthoryear{{Mazzotta}, {Edge} \& {Markevitch}}{{Mazzotta}
  et~al.}{2003}]{Maz2003596}
{Mazzotta} P.,  {Edge} A.~C.,    {Markevitch} M.,  2003, \apj, 596, 190

\bibitem[\protect\citeauthoryear{{Mazzotta} \& {Giacintucci}}{{Mazzotta} \&
  {Giacintucci}}{2008}]{Maz2008675}
{Mazzotta} P.,  {Giacintucci} S.,  2008, \apjl, 675, L9

\bibitem[\protect\citeauthoryear{{McConnell}, {Ma}, {Gebhardt}, {Wright},
  {Murphy}, {Lauer}, {Graham} \& {Richstone}}{{McConnell}
  et~al.}{2011}]{McC2011480}
{McConnell} N.~J.,  {Ma} C.-P.,  {Gebhardt} K.,  {Wright} S.~A.,  {Murphy}
  J.~D.,  {Lauer} T.~R.,  {Graham} J.~R.,    {Richstone} D.~O.,  2011, \nat,
  480, 215

\bibitem[\protect\citeauthoryear{{McDonald}, {Bayliss}, {Benson}, {Foley},
  {Ruel}, {Sullivan}, {Veilleux}, {Aird}, {Ashby}, {Bautz}, {Bazin}, {Bleem},
  {Brodwin}, {Carlstrom}, {Chang}, {Cho}, {Clocchiatti} \& {et al.}}{{McDonald}
  et~al.}{2012}]{McD2012Nat}
{McDonald} M.,  {Bayliss} M.,  {Benson} B.~A.,  {Foley} R.~J.,  {Ruel} J.,
  {Sullivan} P.,  {Veilleux} S.,  {Aird} K.~A.,  {Ashby} M.~L.~N.,  {Bautz} M.,
   {Bazin} G.,  {Bleem} L.~E.,  {Brodwin} M.,  {Carlstrom} J.~E.,  {Chang}
  C.~L.,  {Cho} H.~M.,  {Clocchiatti} A.,    {et al.} 2012, \nat, 488, 349

\bibitem[\protect\citeauthoryear{{McDonald}, {Veilleux}, {Rupke} \&
  {Mushotzky}}{{McDonald} et~al.}{2010}]{McD2010721}
{McDonald} M.,  {Veilleux} S.,  {Rupke} D.~S.~N.,    {Mushotzky} R.,  2010,
  \apj, 721, 1262

\bibitem[\protect\citeauthoryear{{McNamara}, {Nulsen}, {Wise}, {Rafferty},
  {Carilli}, {Sarazin} \& {Blanton}}{{McNamara} et~al.}{2005}]{McN2005433}
{McNamara} B.~R.,  {Nulsen} P.~E.~J.,  {Wise} M.~W.,  {Rafferty} D.~A.,
  {Carilli} C.,  {Sarazin} C.~L.,    {Blanton} E.~L.,  2005, \nat, 433, 45

\bibitem[\protect\citeauthoryear{{McNamara}, {Wise}, {Nulsen}, {David},
  {Sarazin}, {Bautz}, {Markevitch}, {Vikhlinin}, {Forman}, {Jones} \&
  {Harris}}{{McNamara} et~al.}{2000}]{McN2000534}
{McNamara} B.~R.,  {Wise} M.,  {Nulsen} P.~E.~J.,  {David} L.~P.,  {Sarazin}
  C.~L.,  {Bautz} M.,  {Markevitch} M.,  {Vikhlinin} A.,  {Forman} W.~R.,
  {Jones} C.,    {Harris} D.~E.,  2000, \apjl, 534, L135

\bibitem[\protect\citeauthoryear{{Merloni} \& {Heinz}}{{Merloni} \&
  {Heinz}}{2007}]{Mer2007381}
{Merloni} A.,  {Heinz} S.,  2007, \mnras, 381, 589

\bibitem[\protect\citeauthoryear{{Million}, {Allen}, {Werner} \&
  {Taylor}}{{Million} et~al.}{2010}]{Mil2010405}
{Million} E.~T.,  {Allen} S.~W.,  {Werner} N.,    {Taylor} G.~B.,  2010,
  \mnras, 405, 1624

\bibitem[\protect\citeauthoryear{{Novikov} \& {Thorne}}{{Novikov} \&
  {Thorne}}{1973}]{Nov1973}
{Novikov} I.~D.,  {Thorne} K.~S.,  1973, in {Dewitt} C.,  {Dewitt} B.~S.,  eds,
  Black Holes (Les Astres Occlus) {Astrophysics of black holes.}.
pp 343--450

\bibitem[\protect\citeauthoryear{{Nulsen}, {Jones}, {Forman}, {Churazov},
  {McNamara}, {David} \& {Murray}}{{Nulsen} et~al.}{2009}]{Nul2009}
{Nulsen} P.,  {Jones} C.,  {Forman} W.,  {Churazov} E.,  {McNamara} B.,
  {David} L.,    {Murray} S.,  2009, in {Heinz} S.,  {Wilcots} E.,  eds,
  American Institute of Physics Conference Series Vol.~1201 of American
  Institute of Physics Conference Series, {Radio Mode Outbursts in Giant
  Elliptical Galaxies}.
pp 198--201

\bibitem[\protect\citeauthoryear{{Nulsen}, {McNamara}, {Wise} \&
  {David}}{{Nulsen} et~al.}{2005}]{Nul2005628}
{Nulsen} P.~E.~J.,  {McNamara} B.~R.,  {Wise} M.~W.,    {David} L.~P.,  2005,
  \apj, 628, 629

\bibitem[\protect\citeauthoryear{{O'Dea}, {Baum}, {Privon}, {Noel-Storr},
  {Quillen}, {Zufelt}, {Park}, {Edge}, {Russell}, {Fabian}, {Donahue},
  {Sarazin}, {McNamara}, {Bregman} \& {Egami}}{{O'Dea}
  et~al.}{2008}]{ODe2008681}
{O'Dea} C.~P.,  {Baum} S.~A.,  {Privon} G.,  {Noel-Storr} J.,  {Quillen} A.~C.,
   {Zufelt} N.,  {Park} J.,  {Edge} A.,  {Russell} H.,  {Fabian} A.~C.,
  {Donahue} M.,  {Sarazin} C.~L.,  {McNamara} B.,  {Bregman} J.~N.,    {Egami}
  E.,  2008, \apj, 681, 1035

\bibitem[\protect\citeauthoryear{{Ogrean}, {Hatch}, {Simionescu},
  {B{\"o}hringer}, {Br{\"u}ggen}, {Fabian} \& {Werner}}{{Ogrean}
  et~al.}{2010}]{Ogr2010406}
{Ogrean} G.~A.,  {Hatch} N.~A.,  {Simionescu} A.,  {B{\"o}hringer} H.,
  {Br{\"u}ggen} M.,  {Fabian} A.~C.,    {Werner} N.,  2010, \mnras, 406, 354

\bibitem[\protect\citeauthoryear{{Peterson} \& {Fabian}}{{Peterson} \&
  {Fabian}}{2006}]{Pet2006427}
{Peterson} J.~R.,  {Fabian} A.~C.,  2006, \physrep, 427, 1

\bibitem[\protect\citeauthoryear{{Postman}, {Coe}, {Ben{\'{\i}}tez}, {Bradley},
  {Broadhurst}, {Donahue}, {Ford}, {Graur}, {Graves}, {Jouvel}, {Koekemoer},
  {Lemze}, {Medezinski}, {Molino} \& {et al.}}{{Postman}
  et~al.}{2012}]{Pos201225}
{Postman} M.,  {Coe} D.,  {Ben{\'{\i}}tez} N.,  {Bradley} L.,  {Broadhurst} T.,
   {Donahue} M.,  {Ford} H.,  {Graur} O.,  {Graves} G.,  {Jouvel} S.,
  {Koekemoer} A.,  {Lemze} D.,  {Medezinski} E.,  {Molino} A.,    {et al.}
  2012, \apjs, 199, 25

\bibitem[\protect\citeauthoryear{{Quillen}, {Zufelt}, {Park}, {O'Dea}, {Baum},
  {Privon}, {Noel-Storr}, {Edge}, {Russell}, {Fabian}, {Donahue}, {Bregman},
  {McNamara} \& {Sarazin}}{{Quillen} et~al.}{2008}]{Qui2008176}
{Quillen} A.~C.,  {Zufelt} N.,  {Park} J.,  {O'Dea} C.~P.,  {Baum} S.~A.,
  {Privon} G.,  {Noel-Storr} J.,  {Edge} A.,  {Russell} H.,  {Fabian} A.,
  {Donahue} M.,  {Bregman} J.~N.,  {McNamara} B.~R.,    {Sarazin} C.~L.,  2008,
  \apjs, 176, 39

\bibitem[\protect\citeauthoryear{{Rafferty}, {McNamara} \& {Nulsen}}{{Rafferty}
  et~al.}{2008}]{Raf2008687}
{Rafferty} D.~A.,  {McNamara} B.~R.,    {Nulsen} P.~E.~J.,  2008, \apj, 687,
  899

\bibitem[\protect\citeauthoryear{{Rafferty}, {McNamara}, {Nulsen} \&
  {Wise}}{{Rafferty} et~al.}{2006}]{Raf2006652}
{Rafferty} D.~A.,  {McNamara} B.~R.,  {Nulsen} P.~E.~J.,    {Wise} M.~W.,
  2006, \apj, 652, 216

\bibitem[\protect\citeauthoryear{{Randall}, {Forman}, {Giacintucci}, {Nulsen},
  {Sun}, {Jones}, {Churazov}, {David}, {Kraft}, {Donahue}, {Blanton},
  {Simionescu} \& {Werner}}{{Randall} et~al.}{2011}]{Ran2011726}
{Randall} S.~W.,  {Forman} W.~R.,  {Giacintucci} S.,  {Nulsen} P.~E.~J.,  {Sun}
  M.,  {Jones} C.,  {Churazov} E.,  {David} L.~P.,  {Kraft} R.,  {Donahue} M.,
  {Blanton} E.~L.,  {Simionescu} A.,    {Werner} N.,  2011, \apj, 726, 86

\bibitem[\protect\citeauthoryear{{Randall}, {Jones}, {Markevitch}, {Blanton},
  {Nulsen} \& {Forman}}{{Randall} et~al.}{2009}]{Ran2009700}
{Randall} S.~W.,  {Jones} C.,  {Markevitch} M.,  {Blanton} E.~L.,  {Nulsen}
  P.~E.~J.,    {Forman} W.~R.,  2009, \apj, 700, 1404

\bibitem[\protect\citeauthoryear{{Russell}, {McNamara}, {Edge}, {Hogan}, {Main}
  \& {Vantyghem}}{{Russell} et~al.}{2013}]{Rus2013}
{Russell} H.~R.,  {McNamara} B.~R.,  {Edge} A.~C.,  {Hogan} M.~T.,  {Main}
  R.~A.,    {Vantyghem} A.~N.,  2013, \mnras

\bibitem[\protect\citeauthoryear{{Sanders}}{{Sanders}}{2006}]{San2006371CB}
{Sanders} J.~S.,  2006, \mnras, 371, 829

\bibitem[\protect\citeauthoryear{{Sanders} \& {Fabian}}{{Sanders} \&
  {Fabian}}{2006}]{San2006371}
{Sanders} J.~S.,  {Fabian} A.~C.,  2006, \mnras, 371, L65

\bibitem[\protect\citeauthoryear{{Sanders} \& {Fabian}}{{Sanders} \&
  {Fabian}}{2007}]{San2007381}
{Sanders} J.~S.,  {Fabian} A.~C.,  2007, \mnras, 381, 1381

\bibitem[\protect\citeauthoryear{{Sanders}, {Fabian} \& {Taylor}}{{Sanders}
  et~al.}{2005}]{San2005356}
{Sanders} J.~S.,  {Fabian} A.~C.,    {Taylor} G.~B.,  2005, \mnras, 356, 1022

\bibitem[\protect\citeauthoryear{{Shurkin}, {Dunn}, {Gentile}, {Taylor} \&
  {Allen}}{{Shurkin} et~al.}{2008}]{Shu2008383}
{Shurkin} K.,  {Dunn} R.~J.~H.,  {Gentile} G.,  {Taylor} G.~B.,    {Allen}
  S.~W.,  2008, \mnras, 383, 923

\bibitem[\protect\citeauthoryear{{Simionescu}, {Werner}, {B{\"o}hringer},
  {Kaastra}, {Finoguenov}, {Br{\"u}ggen} \& {Nulsen}}{{Simionescu}
  et~al.}{2009}]{Sim2009493}
{Simionescu} A.,  {Werner} N.,  {B{\"o}hringer} H.,  {Kaastra} J.~S.,
  {Finoguenov} A.,  {Br{\"u}ggen} M.,    {Nulsen} P.~E.~J.,  2009, \aap, 493,
  409

\bibitem[\protect\citeauthoryear{{Simionescu}, {Werner}, {Finoguenov},
  {B{\"o}hringer} \& {Br{\"u}ggen}}{{Simionescu} et~al.}{2008}]{Sim2008482}
{Simionescu} A.,  {Werner} N.,  {Finoguenov} A.,  {B{\"o}hringer} H.,
  {Br{\"u}ggen} M.,  2008, \aap, 482, 97

\bibitem[\protect\citeauthoryear{{Sternberg} \& {Soker}}{{Sternberg} \&
  {Soker}}{2008}]{Ste2008384}
{Sternberg} A.,  {Soker} N.,  2008, \mnras, 384, 1327

\bibitem[\protect\citeauthoryear{{Tchekhovskoy}, {Narayan} \&
  {McKinney}}{{Tchekhovskoy} et~al.}{2011}]{Tch2011418}
{Tchekhovskoy} A.,  {Narayan} R.,    {McKinney} J.~C.,  2011, \mnras, 418, L79

\bibitem[\protect\citeauthoryear{{Vasudevan} \& {Fabian}}{{Vasudevan} \&
  {Fabian}}{2007}]{Vas2007381}
{Vasudevan} R.~V.,  {Fabian} A.~C.,  2007, \mnras, 381, 1235

\bibitem[\protect\citeauthoryear{{Werner}, {Oonk}, {Canning}, {Allen},
  {Simionescu}, {Kos}, {van Weeren}, {Edge}, {Fabian}, {von der Linden},
  {Nulsen}, {Reynolds} \& {Ruszkowski}}{{Werner} et~al.}{2013}]{Wer2013767}
{Werner} N.,  {Oonk} J.~B.~R.,  {Canning} R.~E.~A.,  {Allen} S.~W.,
  {Simionescu} A.,  {Kos} J.,  {van Weeren} R.~J.,  {Edge} A.~C.,  {Fabian}
  A.~C.,  {von der Linden} A.,  {Nulsen} P.~E.~J.,  {Reynolds} C.~S.,
  {Ruszkowski} M.,  2013, \apj, 767, 153

\bibitem[\protect\citeauthoryear{{Werner}, {Sun}, {Bagchi}, {Allen}, {Taylor},
  {Sirothia}, {Simionescu}, {Million}, {Jacob} \& {Donahue}}{{Werner}
  et~al.}{2011}]{Wer2011415}
{Werner} N.,  {Sun} M.,  {Bagchi} J.,  {Allen} S.~W.,  {Taylor} G.~B.,
  {Sirothia} S.~K.,  {Simionescu} A.,  {Million} E.~T.,  {Jacob} J.,
  {Donahue} M.,  2011, \mnras, 415, 3369

\bibitem[\protect\citeauthoryear{{White}, {Jones} \& {Forman}}{{White}
  et~al.}{1997}]{Whi1997292}
{White} D.~A.,  {Jones} C.,    {Forman} W.,  1997, \mnras, 292, 419

\bibitem[\protect\citeauthoryear{{ZuHone}, {Markevitch}, {Brunetti} \&
  {Giacintucci}}{{ZuHone} et~al.}{2013}]{ZuH2013762}
{ZuHone} J.~A.,  {Markevitch} M.,  {Brunetti} G.,    {Giacintucci} S.,  2013,
  \apj, 762, 78

\end{thebibliography}

\end{document}